\documentclass[10pt,a4paper,twoside]{article}
\usepackage{epsfig}
\usepackage{baltlat5}
\usepackage{wrapfig}
\begin{document}
\ \ \vspace{-0.5mm}

\setcounter{page}{199} \vspace{-2mm}

\titlehead{Baltic Astronomy, vol.\ts 16, 199--226, 2007.}

\titleb{SEVEN-COLOR VILNIUS PHOTOMETRY OF THE OPEN\hfil\break CLUSTER
NGC\,752}

\begin{authorl}
\authorb{S.~Barta{\v s}i{\= u}t{\. e}}{1,2},
\authorb{V.~Deveikis}{1},
\authorb{V.~Strai{\v z}ys}{2}
and
\authorb{A.~Bogdanovi{\v c}ius}{3}
\end{authorl}

\begin{addressl}
\addressb{1}{Astronomical Observatory of Vilnius University,
\v Ciurlionio 29, Vilnius\\ LT-03100, Lithuania}

\addressb{2}{Institute of Theoretical Physics and Astronomy,
Vilnius University,\\ Go{\v s}tauto 12, Vilnius LT-01108, Lithuania}

\addressb{3}{Department of Physics, Vilnius Gediminas Technical
University,\hfil\break Saul{\. e}tekio al. 11, Vilnius, LT-10223,
Lithuania}

\end{addressl}

\submitb{Received 2006 December 18; revised and accepted 2007 June 1}

\begin{summary}
New photoelectric seven-color observations in the {\it Vilnius}
system are
presented for 65 stars in the region of the open cluster NGC\,752.
Based on individual stars with accurate photometric classifications,
we determine the apparent distance modulus
$(m$$-$$M)_V$\,=\,8.38$\pm$\,0.14 and the mean reddening to the
cluster $E_{Y-V}$=\,0.027$\pm$\,0.010, or
$E_{B-V}$=\,0.034$\pm$\,0.013 (the errors given are the standard
deviations for one star). The mean photometric metallicity for the
main-sequence stars,
[Fe/H]\,=\,$-$0.14$\pm$\,0.03, is found to be slightly lower
than that derived for the red
clump giants, [Fe/H]\,=\,$-$0.08$\pm$\,0.09. This difference
suggests that red giants in later evolutionary phases may not have
zero-age surface values of [Fe/H]. We made use of the least-squares
minimization techniques to fit the Padova theoretical isochrones to
the CMD, when the binary star population is taken into account. By
varying the distance modulus, metallicity and age, the best match
has been found between the seven magnitudes and colors of the
observed stars and those of model binaries, which gives the distance
modulus by 0.2 mag smaller than that derived from individual stars,
i.e., $(m$$-$$M)_V$\,=\,8.18, a closely similar metallicity
([Fe/H]\,=\,$-$0.12), and age of 1.6 Gyr. With these results, the
fraction of photometric binaries among the main-sequence stars is
$\geq$40\%.
\end{summary}


\resthead{Vilnius photometry of the open cluster NGC
752}{S.~Barta{\v s}i{\= u}t{\. e}, V.~Deveikis, V.~Strai{\v z}ys,
A.~Bogdanovi{\v c}ius}

\vskip2mm

\sectionb{1}{INTRODUCTION}

The open cluster NGC\,752 [\,$\alpha(2000)$\,= $01^{\rm h}57^{\rm m}
41^{\rm s}$,\enskip $\delta(2000)$\,= $+37^{\circ}47.1^{\prime}$;\quad
$l$\,=\break \,137.12\degr,\enskip $b$\,=\,--23.26\degr\,],
lying at a distance of $\sim$450 pc (Daniel et al. 1994), continues
to receive much attention because it is one of the very few clusters
in the solar neighborhood with an age between 1 and 2 Gyr,
critically important to tests of stellar evolution. For clusters in
such an age-range the theory predicts the effects of convective core
overshooting (for stars leaving the main sequence), an extended
red-giant branch (RGB) populated by stars burning hydrogen in a
shell, a clump of red giants which have started their core
helium-burning phase, and a sparsely populated asymptotic giant
branch.

According to a summary of observational evidence available at the
present time, NGC\,752 contains a relatively small number of stars
and is of low central concentration, with a corona extending to
about 100$^\prime$ from the cluster center\break (Mermilliod et al.
1998). The distinguishing feature of this cluster is the morphology
of its color-magnitude diagram (CMD). The greatest number of stars
is found in the region of the turnoff, which occurs for early F-type
stars, and along the upper part of the unevolved main sequence,
whereas the lower part of the main sequence (i.e., the domain of
stars of G-type and later) appears to be very sparsely populated. In
such an intermediate-age cluster this may be an indication of the
dynamic escape of low mass stars from the cluster. The rich
population of binaries and the presence stars with rapid rotation
(for types earlier than F5) make the main sequence and its turnoff
point difficult to define. Containing only a handful of red giants,
this cluster does not display a clearly defined RGB either. All of
the 15 known red giants are located within, or very close to, the
region of the CMD, which can be identified as the locus of the red
clump stars. As it has been pointed out by Mermilliod et al. (1998),
the red clump is observed to present a faint extension slightly to
the blue of its main concentration, which is not accounted for by
theoretical isochrones (see, also, Girardi et al. 2000). Also, the
cluster contains one blue straggler of type A0, which was reported
to be a spectroscopic binary with a long orbital period (Latham et
al., private communication to van den Berg \& Verbunt 2001).

Since the first proper-motion-based identification of the cluster
members by Ebbighausen (1939), complete down to $m_{\rm
pg}$$\approx$\,12.2 within a diameter of 50$^\prime$, the census of
probable members down to fainter magnitudes has been extended
through three modern proper-motion surveys. One was that of Cannon
(1968), in which the virtual absence of fainter members was noted
between $B$ = 15 mag and the proper-motion plate limit at $B$ = 17 mag.
A more recent survey by Francic (1989), complete down to
$V$\,$\approx$\,14 mag, also has shown the deficiency of faint members
of the unevolved main sequence. The third and the most extensive and
reliable census of the proper-motion members was accomplished by
Platais (1991) in a much larger area outlined by a 110$^\prime$
diameter circle around the center of the cluster. In that study,
complete down to $m_{\rm pg}$=\,15.0, the total number of stars
identified to be probable members reached 130, with the luminosity
function showing again a deficiency of faint members of the main
sequence. According to the above three photographic studies, as well
as that by Rohlfs \& Van{\'y}sek (1962), the main sequence extends
down to about $V$ = 14.4 mag, contrary to some earlier statements made
from photoelectric photometry that the sequence breaks off at about
12th magnitude (e.g., Arp 1962; Eggen 1963).

More recently, probable members were thoroughly identified and
analyzed in a comprehensive study by Daniel et al. (1994, hereafter
 DLMT), based on all available photometric and
proper-motion data and radial-velocity observations. A long-term
radial-velocity study of 30 red giants within $2^\circ$ from the
cluster center was carried out by Mermilliod et al. (1998) who
confirmed the membership of 15 such stars. Therefore, the existing
proper-motion and radial velocity data now give a more-or-less
complete census of all probable members within the observed cluster
diameter ($\sim$\,110$^\prime$) and to the magnitude limit
$V$\,$\sim$\,14.5.

Over the years, a number of photoelectric studies of NGC\,752 have
been done in different photometric systems ({\it UBV}, {\it UBViyz},
{\it uvby}$\beta$, {\it DDO}, {\em Washington}, {\em Geneva} and {\em
Vilnius}), which gave results on the main cluster parameters
generally in agreement. From combined photometry in the literature
up to 1993, DLMT obtained the weighted mean values
$E_{B-V}$=\,0.035$\pm$\,0.005, $(m$$-$$M)_0$=\,8.25$\pm$\,0.10 and
[Fe/H]\,=\,$-$0.15\,$\pm$\,0.05 (uncertainties quoted are probable
errors). With these cluster parameters they arrived at an age
estimate of 1.7$\pm$\,0.1 Gyr by comparison with the overshoot
models of Meynet et al. (1993). Age dating based on the more recent
model isochrones with overshooting gives the values which are likely
to converge toward 1.6 Gyr (c.f. Kozhurina-Platais et al. 1997; Pols
et al. 1998; Anthony-Twarog \& Twarog 2004, 2006). Recently,
Anthony-Twarog \& Twarog (2006) have published new photoelectric
photometry on the extended Str{\"o}mgren {\it uvbyCa} system for 7 red
giants and 21 main-sequence stars of the cluster. From 10
single-star members they found the mean metallicity
[Fe/H]\,=\,$-$0.06$\pm$\,0.03 (s.e.m.), on a scale where the Hyades
have [Fe/H]\,=\,$+$0.12.

Spectroscopic observations of the cluster members have added
independent measures of the cluster metallicity. From
medium-resolution spectra of nine red giants Friel \& Janes (1993)
obtained [Fe/H]\,=\,$-$0.16$\pm$\,0.05 (s.d.), or
[Fe/H]\,=\,$-$\,0.18$\pm$\break 0.04 according to the updated
abundance calibration (Friel et al. 2002). However, the available
few estimates based on high-resolution spectra range from slightly
subsolar to solar values: from eight narrow-lined F-type
main-sequence stars Hobbs \& Thorburn (1992) arrived at a value of
[Fe/H]\,=\,$-$0.09$\pm$\,0.05, whereas, more recently, Sestito et
al. (2004) obtained $+$0.01$\pm$\,0.04 for 18 main-sequence F8--G8
stars, based on comparison with the Hyades for which
[Fe/H]\,=\,$+$0.12. More subtle features in the spectra of the
cluster stars, such as absorption lines of Li, were also
investigated (Hobbs \& Pilachowski 1986; Pilachowski et al. 1988;
Gilroy 1989; Sestito et al. 2004).

Application of intermediate-band seven-color {\em Vilnius}
photometry can provide precise estimates of absolute magnitudes,
foreground reddening and metallicities for individual stars of
nearly all spectral types, leading to improved determinations of the
main parameters of Galactic clusters. Some evolutionary phenomena
such as mass loss and internal mixing in evolving stars can also be
discernible from accurate multicolor photometry. In the paper of
Barta\v si\= ut\. e \& Tautvai{\v s}ien{\. e} (2004), a photometric
analysis by means of {\em Vilnius} system of the red giants in the
open cluster NGC\,7789, which is about the same age as NGC\,752,
revealed that the red clump giants have a slightly higher
photometric metallicity than the ascending giant branch stars. This
received an explanation in terms of such phenomena as mixing which
can change the intensities of carbon and nitrogen molecular bands
many of which fall in the passbands of the {\em Vilnius} system. It
is therefore of interest to see whether that interpretation is
supported by the evidence from analysis of photometric metallicities
of the red clump giants and unevolved stars in NGC\,752.

In the {\em Vilnius} system, NGC\,752 was first observed by
Dz{\`e}rv{\`i}tis \& Paupers (1993, hereafter DP).
From photoelectric seven-color data for 89 stars in the field of the
cluster they derived the mean cluster reddening
$E_{B-V}$\,=\,0.025\,$\pm$\break 0.025 and the distance modulus
$(m$$-$$M)_0$\,=\,8.15\,$\pm$\,0.15 (by main-sequence fitting and
adopting for the Hyades distance modulus 3.35 mag). However they
made no attempt to determine metallicity. The majority of the stars
were observed by DP on only one or two night(s). To obtain a
reliable estimate of the cluster metallicity, good photometry,
exclusion of binaries and a firmly established value of reddening
are required. Variations of 0.02 or 0.03 mag in color have often
little effect on the derived modulus, however for [Fe/H]
determinations these introduce a noticeable error. Therefore, for
the purpose of an analysis of the cluster metallicity, we performed
new seven-color observations in the {\em Vilnius} system. While our
eventual goal is to obtain {\em Vilnius} CCD observations of a
complete sample of stars down to the fainter members of the cluster,
we made during our photoelectric part of the program as many
observations per star as would be desired for photoelectric
standards in the future CCD reductions.

In this paper we present the new photoelectric photometry in the
{\em Vilnius} system of 65 stars in the region of NGC\,752 and
determine the main parameters of this cluster, with a primary
emphasis put on the impact of the binary-star population on both the
parameters derived and the CMD fit. The acquisition of observational
data is described in Section~2. In Section~3 we discuss briefly the
methods used for the photometric determination of stellar parameters
and present them for individual stars. In Section~4 we determine the
basic parameters of the cluster: the interstellar reddening and
distance modulus (\S\,4.1), metallicity (\S\,4.2) and, finally, in
\S\,4.3 provide the CMD fit to the theoretical isochrones and
discuss the binary-star population in the cluster. The results are
summarized in Section~5.

\vskip2mm

\sectionb{2}{OBSERVATIONS}

Photoelectric observations in the intermediate-band {\em Vilnius}
system were made in the fall observing run of 1997 using a
single-channel photometer attached to the 1-meter telescope at
Maidanak Observatory (altitude 2550 m) in Uzbekistan. The photometer
operated with a multi-alkali FEU-79 phototube and a standard set of
{\em UPXYZVS} filters to establish the {\em Vilnius} system.
Integration times were set to achieve counts of at least 10\,000
above the sky through each filter. Each star was observed on at
least three different nights. To allow the determination of
atmospheric extinction coefficients by the Nikonov method (described
by Strai{\v z}ys 1992 and Zdanavi{\v c}ius 1996), a pair of standard
stars, HD\,4568 (F8\,V) and HD\,199598 (G0\,V), separated by a wide
range of air mass, were observed in 1$-$1.5 hour intervals during
each photometric night. Nikonov's method takes into account the
variations of atmospheric extinction during the night and the
dependence of extinction coefficients on spectral type and
interstellar reddening. The transformation of the magnitudes $V$ and
color indices in the instrumental system to the standard {\em
Vilnius} system was made through observations of 28 standard stars,
mostly in the Cygnus Standard Region (Zdanavi{\v c}ius \& {\v
C}ernien{\. e} 1985). The standard errors of the transformation
equations were 0.004 for magnitude $V$ and color index $U$$-$$V$,
0.003 for $P$$-$$V$, $Y$$-$$V$ and $V$$-$$S$, and 0.002 for the
remaining color indices $X$$-$$V$ and $Z$$-$$V$. A total of 65 stars
have been measured in the field of the cluster.

The photometry is presented in Table~1. The first column lists the
Heinemann (1926) number, the second to eighth, the $V$-magnitude and
the color indices of the {\em Vilnius} system, and the ninth column
gives the number of observations on each star. Of the 65 stars
observed, three are known as nonmembers (H\,186, H\,220 and H\,226)
from proper-motion and/or radial-velocity studies in the literature.
In the last column these three nonmember stars are marked with the
designation ``NM". The $V$ magnitudes of the two eclipsing
variables, H\,219 and H\,235, exhibiting significant magnitude
variations, are marked with the letter ``v".

The internal precision of photometry in Table~1 is typically better
than 0.010 mag and appears to be not very magnitude-dependent. The
internal mean errors are 0.008 mag for $U$$-$$V$, 0.007 mag for $V$
and the color indices $P$$-$$V$ and $V$$-$$S$, 0.006 mag for
$Z$$-$$V$, and 0.005 mag for $X$$-$$V$ and $Y$$-$$V$.

~
\\[-36pt]
\begin{center}
\vbox{\footnotesize \tabcolsep=5pt
\begin{tabular}{rrrrrrrrcc}
\multicolumn{10}{c}{\parbox{108mm}{\baselineskip=12pt {\normbf \ \
Table~1.}{\norm\ {\it Vilnius} photometry in the field of the open
cluster NGC\,752.}}}
\\[2pt]
\tablerule
{Heinema\rlap{nn}} & & & & & & & & & \\
\multicolumn{1}{r}{No.} & \multicolumn{1}{c}{$V~~$} &
\multicolumn{1}{c}{$U$$-$$V$} & \multicolumn{1}{c}{$P$$-$$V$} &
\multicolumn{1}{c}{$X$$-$$V$} & \multicolumn{1}{c}{$Y$$-$$V$} &
\multicolumn{1}{c}{$Z$$-$$V$} &
\multicolumn{1}{c}{$V$$-$$S$} & \multicolumn{1}{c}{$n$} & \multicolumn{1}{c}{Rem.}\\
\tablerule
         1 &  9.505 & 3.443 & 2.863 & 1.961 & 0.755 & 0.294 & 0.704 & 3 & \\
        10 & 10.141 & 2.274 & 1.696 & 1.046 & 0.457 & 0.160 & 0.424 & 3 & \\
        11 &  9.296 & 3.399 & 2.837 & 1.922 & 0.736 & 0.286 & 0.700 & 3 & \\
        12 &  9.948 & 2.313 & 1.759 & 1.121 & 0.484 & 0.177 & 0.473 & 3 & \\
        41 &  9.821 & 2.382 & 1.797 & 1.157 & 0.495 & 0.182 & 0.474 & 3 & \\[5pt]
        55 & 11.433 & 2.229 & 1.702 & 1.098 & 0.484 & 0.185 & 0.419 & 3 & \\
        58 & 10.484 & 2.247 & 1.691 & 1.027 & 0.448 & 0.157 & 0.412 & 3 & \\
        61 & 10.044 & 2.272 & 1.677 & 1.014 & 0.438 & 0.153 & 0.413 & 3 & \\
        62 & 11.213 & 2.188 & 1.648 & 1.048 & 0.452 & 0.166 & 0.421 & 3 & \\
        63 & 11.051 & 2.216 & 1.706 & 1.122 & 0.477 & 0.168 & 0.464 & 3 & \\[5pt]
        64 & 10.540 & 2.272 & 1.671 & 0.984 & 0.416 & 0.146 & 0.384 & 3 & \\
        66 & 10.923 & 2.190 & 1.684 & 1.091 & 0.474 & 0.172 & 0.446 & 3 & \\
        69 & 10.054 & 2.440 & 1.838 & 1.188 & 0.496 & 0.175 & 0.475 & 3 & \\
        74 & 10.735 & 2.259 & 1.700 & 1.027 & 0.436 & 0.163 & 0.421 & 3 & \\
        75 &  8.977 & 3.545 & 2.966 & 2.025 & 0.775 & 0.302 & 0.726 & 3 & \\[5pt]
        77 &  9.393 & 3.628 & 3.048 & 2.069 & 0.789 & 0.311 & 0.752 & 4 & \\
        88 & 11.735 & 2.257 & 1.758 & 1.176 & 0.514 & 0.208 & 0.470 & 3 & \\
        96 & 10.376 & 2.299 & 1.680 & 0.995 & 0.414 & 0.148 & 0.402 & 3 & \\
       105 & 10.263 & 2.316 & 1.712 & 1.038 & 0.446 & 0.156 & 0.432 & 4 & \\
       106 & 10.512 & 2.236 & 1.661 & 0.978 & 0.426 & 0.162 & 0.396 & 4 & \\[5pt]
       108 &  9.166 & 2.383 & 1.770 & 1.101 & 0.473 & 0.171 & 0.454 & 3 & \\
       110 &  8.968 & 3.110 & 2.536 & 1.717 & 0.691 & 0.272 & 0.669 & 3 & \\
       117 & 10.249 & 2.359 & 1.743 & 1.087 & 0.469 & 0.173 & 0.439 & 3 & \\
       123 & 11.182 & 2.199 & 1.646 & 1.028 & 0.443 & 0.165 & 0.403 & 3 & \\
       126 & 10.102 & 2.314 & 1.720 & 1.060 & 0.464 & 0.163 & 0.443 & 3 & \\[5pt]
       135 & 11.215 & 2.245 & 1.719 & 1.097 & 0.484 & 0.190 & 0.436 & 3 & \\
       137 &  8.927 & 3.588 & 3.007 & 2.043 & 0.781 & 0.304 & 0.732 & 3 & \\
       139 & 11.750 & 2.216 & 1.727 & 1.121 & 0.497 & 0.173 & 0.440 & 3 & \\
       140 & 11.785 & 2.332 & 1.876 & 1.288 & 0.550 & 0.215 & 0.493 & 3 & \\
       171 & 10.198 & 2.329 & 1.733 & 1.069 & 0.457 & 0.163 & 0.457 & 3 & \\[5pt]
       177 & 10.164 & 2.370 & 1.774 & 1.141 & 0.492 & 0.178 & 0.468 & 4 & \\
       185 & 12.241 & 2.281 & 1.786 & 1.200 & 0.520 & 0.195 & 0.502 & 3 & \\
       186 & 10.225 & 3.278 & 2.753 & 1.875 & 0.729 & 0.295 & 0.715 & 5 & NM\\
       187 & 10.433 & 2.222 & 1.690 & 1.051 & 0.451 & 0.167 & 0.435 & 3 & \\
       189 & 11.274 & 2.199 & 1.686 & 1.068 & 0.466 & 0.166 & 0.428 & 3 & \\[5pt]
       192 & 10.742 & 2.258 & 1.685 & 1.026 & 0.439 & 0.158 & 0.420 & 3 & \\
       193 & 10.196 & 2.350 & 1.720 & 1.012 & 0.430 & 0.168 & 0.393 & 3 & \\
       196 & 10.251 & 2.321 & 1.715 & 1.055 & 0.459 & 0.170 & 0.431 & 4 & \\
       197 & 11.596 & 2.205 & 1.684 & 1.100 & 0.472 & 0.176 & 0.439 & 5 & \\
       205 &  9.901 & 2.242 & 1.699 & 1.059 & 0.459 & 0.166 & 0.440 & 6 & \\[5pt]
       206 & 10.032 & 2.372 & 1.776 & 1.125 & 0.486 & 0.172 & 0.470 & 6 & \\
       209 &  9.730 & 1.901 & 1.353 & 0.549 & 0.212 & 0.072 & 0.142 & 3 & \\
       213 &  9.044 & 3.530 & 2.939 & 2.011 & 0.772 & 0.296 & 0.734 & 6 & \\
       214 & 10.459 & 2.301 & 1.698 & 0.996 & 0.425 & 0.153 & 0.399 & 7 & \\
       217 & 10.432 & 2.330 & 1.723 & 1.060 & 0.458 & 0.167 & 0.434 & 6 & \\[5pt]
       218 & 10.071 & 2.336 & 1.753 & 1.111 & 0.470 & 0.177 & 0.447 & 3 & \\
       219 & 10.504\rlap{v} & 2.240 & 1.673 & 1.037 & 0.440 & 0.164 & 0.420 & 3 & \\
       220 &  9.608 & 3.514 & 3.005 & 2.015 & 0.747 & 0.356 & 0.746 & 3 & NM \\
       222 & 10.968 & 2.216 & 1.666 & 1.003 & 0.432 & 0.155 & 0.415 & 6 & \\
       226 & 11.710 & 3.133 & 2.671 & 1.839 & 0.705 & 0.336 & 0.729 & 4 & NM \\
\tablerule
\end{tabular}
}
\end{center}

\begin{center}
\vbox{\footnotesize \tabcolsep=5pt
\begin{tabular}{rrrrrrrrcc}
\multicolumn{10}{c}{\parbox{108mm}{\baselineskip=12pt
{\normbf \ \ Table~1.}{\norm\ (continued)}}}\\
[3pt]
\tablerule
{Heinema\rlap{nn}} & & & & & & & & & \\
\multicolumn{1}{r}{No.} & \multicolumn{1}{c}{$V~~$} &
\multicolumn{1}{c}{$U$$-$$V$} & \multicolumn{1}{c}{$P$$-$$V$} &
\multicolumn{1}{c}{$X$$-$$V$} & \multicolumn{1}{c}{$Y$$-$$V$} &
\multicolumn{1}{c}{$Z$$-$$V$} &
\multicolumn{1}{c}{$V$$-$$S$} & \multicolumn{1}{c}{$n$} & \multicolumn{1}{c}{Rem.}\\
\tablerule
       232 & 11.646 & 2.185 & 1.682 & 1.090 & 0.469 & 0.177 & 0.456 & 4 & \\
       234 & 10.683 & 2.251 & 1.708 & 1.067 & 0.455 & 0.168 & 0.442 & 6 & \\
       235 & 11.403\rlap{v} & 2.224 & 1.722 & 1.147 & 0.497 & 0.181 & 0.477 & 5 & \\
       237 & 12.333 & 2.395 & 1.904 & 1.284 & 0.551 & 0.193 & 0.552 & 3 & \\
       238 &  9.968 & 2.311 & 1.733 & 1.085 & 0.474 & 0.169 & 0.441 & 4 & \\[5pt]
       254 & 10.913 & 2.229 & 1.666 & 0.991 & 0.428 & 0.163 & 0.399 & 3 & \\
       259 & 11.382 & 2.180 & 1.654 & 1.059 & 0.471 & 0.181 & 0.421 & 3 & \\
       261 & 11.172 & 2.209 & 1.729 & 1.137 & 0.486 & 0.176 & 0.469 & 3 & \\
       263 & 10.950 & 2.197 & 1.638 & 0.972 & 0.413 & 0.144 & 0.413 & 3 & \\
       266 & 11.212 & 2.182 & 1.655 & 1.050 & 0.461 & 0.169 & 0.420 & 3 & \\[5pt]
       295 &  9.311 & 3.396 & 2.829 & 1.929 & 0.748 & 0.289 & 0.708 & 3 & \\
       300 &  9.595 & 2.291 & 1.688 & 1.040 & 0.452 & 0.168 & 0.423 & 3 & \\
       302 & 11.517 & 2.175 & 1.666 & 1.054 & 0.447 & 0.155 & 0.436 & 3 & \\
       304 & 11.875 & 2.201 & 1.723 & 1.120 & 0.483 & 0.182 & 0.470 & 3 & \\
       311 &  9.070 & 3.654 & 3.046 & 2.082 & 0.792 & 0.307 & 0.747 & 3 & \\
\tablerule
\end{tabular}
\begin{list}{}{}
\item[NM:] nonmember by proper-motion and/or radial-velocity criteria
(Pilachowski et al. 1988; Platais 1991; Daniel et al. 1994).
\end{list}
}
\end{center}

A comparison with previous {\em Vilnius} photometry by DP for 59
stars in common shows no systematic discordance between the colors
in both samples. We note that for star H\,110 the color index
$V$$-$$S$ given by DP in their Table~1 is clearly erroneous, most
probably a misprint. Excluding this star we find the mean residuals
to be in the range from 0.001 mag (the smallest difference; for
$Z$$-$$V$) to 0.008 mag (the largest difference; for $X$$-$$Y$),
with residual dispersion varying from $\pm$\,0.015 (for $Y$$-$$Z$)
to 0.027 (for $U$$-$$P$). However our magnitudes $V$ compare much
less favorably with DP observations (middle panel of Figure~1). For
two stars, H\,55 and H\,189, the differences in $V$ amongst the two
works are larger than 0.5 mag. Excluding these two deviants and two
more stars with $\Delta\,V$ exceeding 0.15 mag, the average residual
in $V$ and the standard deviation are reduced to $-$0.015 and
$\pm$\,0.032 mag, respectively, with the magnitudes of our
observations being systematically fainter.

Since $V$ magnitudes of the {\em Vilnius} system are practically
identical to those of the {\it UBV} system, a direct comparison can be
made with several other photoelectric studies in NGC\,752. Figure~1
shows few such examples with the largest number of stars found in
common. The mean differences between the values in the literature
and our values range from $-$0.023 to $+$0.024 mag. The residual
dispersion is found to be generally small in either case:
0.008--0.020 mag. However, from comparison with the magnitudes $V$
of {\it Geneva} photometry (Rufener 1988) or those from recent
{\it uvbyCa} observations by Anthony-Twarog \& Twarog (2006), we see a
clear slope in $\Delta$$V$ vs. $V$, with our magnitudes tended to
become progressively brighter with increasing $V$. No such slope is
found from comparison with first {\it UBV} measurements in NGC\,752 made
by Johnson (1953), except for the systematic difference of the order
of 0.02 mag (top panel), long noted in the literature (e.g. Eggen
1963; DLMT). The very good agreement was found with $V$ magnitudes
from Joner \& Taylor (1995) in an internally consistent {\it
Str{\"o}mgren} system: for 13 stars in common we find the mean
residual $-$0.002 mag, with a standard deviation of $\pm$\,0.008
mag.

\begin{figure}[h!]
\centerline{\psfig{figure=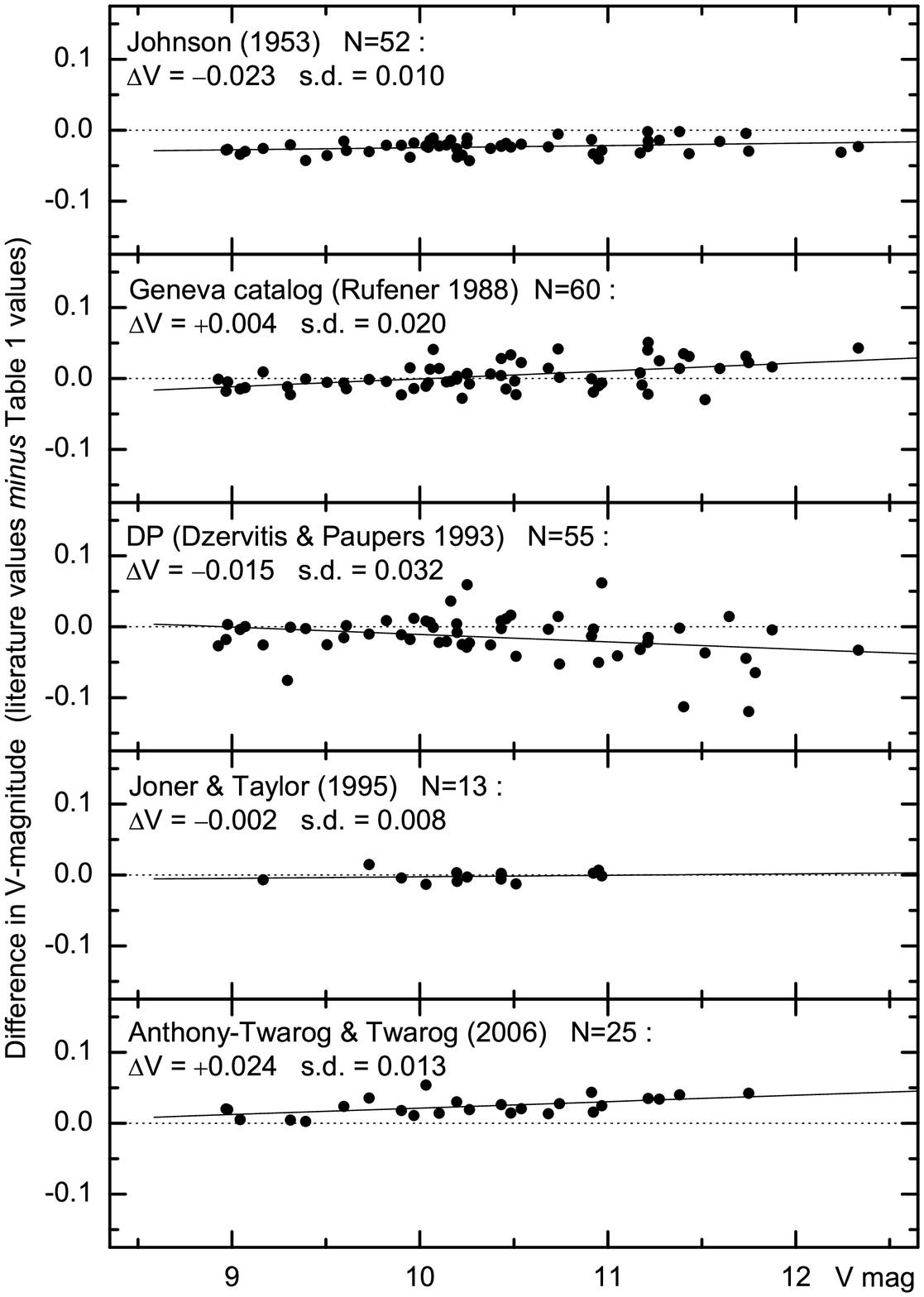,width=96truemm,angle=0,clip=}}
\captionb{1}{Residuals in {\itl V} between the values in the
literature and our values, plotted against magnitude for NGC\,752
stars in common. Solid line is a linear fit to the points.}
\end{figure}

\sectionb{3}{DETERMINATION OF STELLAR PARAMETERS}

Photometric classification in terms of spectral type, absolute
magnitude $M_V$, and color excess $E_{Y-V}$ due to interstellar
reddening was performed by the method analogous to ``Geneva
photometric boxes" developed by Geneva photometrists (e.g., Nicolet
1981a). The principle of the method is to find from an extended
database a set of standard stars (``box") having the same
reddening-free parameters $Q$ and hence the same physical parameters
as the observed star and then to ascribe to the latter the
appropriate ``box" parameters: spectral type, $M_V$, [Fe/H] and
intrinsic color index $(Y$$-$$V)_0$ which is then subtracted from
the observed $Y$$-$$V$ to obtain the reddening $E_{Y-V}$. The method
matches a series of different reddening-free $Q$ parameters (up to
14) which are defined as
  \begin{equation}
           Q_{1234} = (m_1-m_2) - (E_{12}/E_{34})(m_3-m_4)\,,
   \end{equation}
where $E_{12}/E_{34}$ is the color excess ratio, and extracts for
each star a set of standards for which
   \begin{equation}
           \Delta Q = \sqrt {{\sum\nolimits_{i=1}^n (\Delta
           Q_i)^2}/n}
    \end{equation}
is found to be a minimum. Here, $\Delta Q_i$ are the differences
between the corresponding $Q$-parameters of the observed star and
the standard star, and $n$ is the number of $Q$-parameters involved.
If $\Delta Q$ values in a given ``box" do not exceed 0.02, the
derived stellar parameters should be accurate to within typical
errors of classification in the {\em Vilnius} system: about $\pm$1
spectral subclass, $\pm$0.03 mag in $E_{Y-V}$, $\pm$0.5 mag in
$M_V$, and $\pm$0.2 dex in [Fe/H]. The database of standard stars
was compiled of $\sim$\,4400 stars having both precise {\em Vilnius}
photometry and Hipparcos parallaxes and covering nearly the entire
spectral, luminosity and metal abundance range. Various types of
peculiar and binary stars are also included in the database. The
method works remarkably well, with accuracy of classification
dependent primarily on the errors in photometry.

We have used in the present work five different parameters $Q$
($Q_{UPYV}$, $Q_{PXYV}$, $Q_{XYV}$, $Q_{ZVYV}$ and $Q_{VSYV}$) and a
set of five standard stars in each extracted photometric ``box" for
averaging their physical parameters. $Q$-parameters of the stars
observed in NGC\,752 were calculated using the normal extinction law
(Table~64 of Strai\v{z}ys 1992).

The individual values of interstellar reddening $E_{Y-V}$, obtained
in the above way, were averaged to derive the resulting mean cluster
reddening (\S\,4.1), and this value was next used for the
determination of all intrinsic color-indices of the {\em Vilnius}
system. Once the latter are known, the ``photometric box" method can
be used in the manner just described to match the six intrinsic
color indices, $(CI)_{0i}$, instead of five $Q$-parameters. In this
case, we extract the ``box" for which
  \begin{equation}
           \Delta (CI)_0 = \sqrt {{\sum\nolimits_{i=1}^n (\Delta
           (CI)_{0i})^2}/n}
    \end{equation}
is a minimum. In practice, the $Q$- and $(CI)_{0i}$-based ``boxes"
give self-consistent results.

We note that metallicities obtained in this way, i.e., taken as
average from five closest standard stars in the ``box", will be
given further in this work a lower priority than those obtained
directly from the calibrations of {\em Vilnius} indices. To
determine the latter estimates of [Fe/H], we used empirical
calibrations of Bartkevi{\v c}ius \& Sperauskas (1983) that
incorporate a line-blanketing effect on the color indices $P$$-$$X$,
$X$$-$$Y$ and $P$$-$$Y$. The calibrations apply to dwarf stars of
spectral types F--K (0.35\,$\leq$\,$Y$$-$$Y$\,$\leq$\,0.80), G--K
subgiants (0.55\,$\leq$\,$Y$$-$$Y$\,$\leq$\,0.80), and giants of
spectral types G to M (0.55\,$\leq$\,$Y$$-$$Y$\,$\leq$\,1.20) over
the entire range of metallicities.

The resulting photometric parameters are given in Table~2. The first
column lists the star number, and the second to seventh give the
photometric spectral type, intrinsic color index ($Y$$-$$V$)$_0$,
color excess $E_{Y-V}$, absolute magnitude $M_V$, apparent distance
modulus $(m$$-$$M)_V$, and metallicity [Fe/H]. Here, the temperature
sensitive indices ($Y$$-$$V$)$_0$ are inferred from the observed
$Y$$-$$V$ by subtracting the mean\break
\\[-36pt]
\begin{center}
\vbox{\footnotesize\tabcolsep=5pt
\begin{tabular}{rlrrrrrrllll}
\multicolumn{12}{c}{\parbox{120mm}{\baselineskip=12pt {\normbf\ \
Table~2.}{\norm\ Photometric parameters for stars in the field of
NGC\,752.}}}\\[2pt]
\tablerule \\[-6pt]
No. & Sp & \llap{($Y$$-$$V$)}$_0$ & $E$$_{Y-V}$ & $M_V\,$ & \llap{($m$}{\kern-0.05em}--{\kern-0.05em}$M$\rlap{)$_V$} & ~[Fe/H\rlap{]} & $\Delta Q$~ & \multispan{4}~Notes: RV{\kern-0.07em}, SB{\kern-0.07em}, ROT{\kern-0.07em}, etc\rlap{.} \\
\tablerule \\[-6pt]
  1 & G8\,III   & 0.728 & 0.025 & 0.82 & 8.69 & $-$0.10 & 0.007 & R\rlap{V} & & & \\
 10 & F2.5\,IV  & 0.430 & 0.027 & 2.27 & 7.87 & $-${\em 0.17} & 0.008 & & & & \\
 11 & G8\,III   & 0.709 & 0.009 & 0.76 & 8.54 & +0.11 & 0.008 & R\rlap{V} & & & \\
 12 & F4\,IV    & 0.457 & 0.024 & 2.56 & 7.39 & $-${\em 0.12} & 0.007 & R\rlap{V} & SB2~ORB & & \\
 41 & F6\,IV    & 0.468 & 0.031 & 1.93 & 7.89 & $-${\em 0.07} & 0.009 & R\rlap{V} & & RO\rlap{T} & \\[5pt]
 55 & F4\,IV--V & 0.457 & 0.048 & 3.04 & 8.39 & $-$0.06 & 0.010 & R\rlap{V} & SB1 & & \\
 58 & F2\,IV--V & 0.421 & 0.018 & 2.47 & 8.01 & $-${\em 0.09} & 0.010 & & & & \rlap{$\ast$} \\
 61 & F2\,IV    & 0.411 & 0.027 & 2.25 & 7.79 & $-${\em 0.07} & 0.012 & R\rlap{V} & SB2~ORB & & \\
 62 & F3\,V     & 0.425 & 0.015 & 2.87 & 8.34 & $-$0.19 & 0.010 & R\rlap{V} & SB1 & & \\
 63 & F5\,V     & 0.450 & 0.016 & 3.25 & 7.80 & $-$0.11 & 0.007 & R\rlap{V} & SB2 & & \\[5pt]
 64 & F1\,IV    & 0.389 & 0.021 & 2.32 & 8.22 & {\em +0.05} & 0.010 & & & & \\
 66 & F4\,IV--V & 0.447 & 0.022 & 3.11 & 7.81 & $-$0.15 & 0.008 & R\rlap{V} & & & \\
 69 & F6\,III--IV & 0.469 & 0.026 & 1.83 & 8.22 & $-${\em 0.15} & 0.011 & & & & \\
 74 & F2\,IV\,bin. & 0.409 & 0.041 & 2.42 & 8.31 & {\em +0.01} & 0.007 & R\rlap{V} & SB1 & & \\
 75 & G9\,III   & 0.748 & 0.015 & 0.59 & 8.39 & $-$0.06 & 0.007 & R\rlap{V} & SB1~ORB & & \\[5pt]
 77 & G9\,III   & 0.762 & 0.028 & 0.92 & 8.47 & $-$0.09 & 0.007 & R\rlap{V} & & & \rlap{$\ast$} \\
 88 & F6\,V     & 0.487 & 0.036 & 3.43 & 8.30 & $-$0.10 & 0.010 &  R\rlap{V} & SB1 & & \\
 96 & F1\,IV    & 0.387 & 0.031 & 2.27 & 8.11 & {\em +0.03} & 0.010 & & & & \\
105 & F2\,IV    & 0.419 & 0.038 & 2.05 & 8.21 & $-${\em 0.04} & 0.010 & & & & \\
106 & F2\,IV    & 0.399 & 0.029 & 2.48 & 8.03 & $-${\em 0.18} & 0.009 & & & & \\[5pt]
108 & F3\,IV    & 0.446 & 0.033 & 1.86 & 7.31 & {\em +0.06} & 0.009 & R\rlap{V} & & RO\rlap{T} & \\
110 & G6\,III   & 0.664 & 0.021 & 0.80 & 8.17 & $-$0.11\rlap{:} & 0.014 &  R\rlap{V} & SB1~ORB & \\
117 & F3\,IV    & 0.442 & 0.045 & 1.99 & 8.26 & $-${\em 0.01} & 0.010 & & & RO\rlap{T} & \\
123 & F2\,V     & 0.416 & 0.025 & 2.78 & 8.40 & $-$0.14 & 0.010 &  & & & \\
126 & F3\,IV    & 0.437 & 0.034 & 2.05 & 8.05 & $-${\em 0.12} & 0.009 & & & & \\[5pt]
135 & F4\,IV--V & 0.457 & 0.038 & 2.83 & 8.38 & $-$0.04\rlap{:} & 0.009 & & & & \\
137 & G9\,III   & 0.754 & 0.017 & 0.69 & 8.24 & $-$0.08 & 0.007 &  R\rlap{V} & & & \\
139 & F5\,V     & 0.470 & 0.037 & 3.38 & 8.37 & $-$0.13\rlap{:} & 0.012 & R\rlap{V} & & RO\rlap{T} & \\
140 & F9\,V\,bin. & 0.523 & 0.028 & 3.90 & 7.88 & $-$0.13 & 0.015 & & & & \\
171 & F2\,IV    & 0.430 & 0.017 & 1.99 & 8.21 & $-${\em 0.02} & 0.009 & & & & \\[5pt]
177 & F6\,IV    & 0.465 & 0.048 & 1.96 & 8.20 & $-${\em 0.14} & 0.009 & & & RO\rlap{T} & \\
185 & F7\,V     & 0.493 & 0.034 & 3.47 & 8.77 & $-$0.15 & 0.007 & R\rlap{V} & & & \\
186 & G9\,IV    & 0.680 & 0.049 & 3.40 & 6.83 & $-$0.14 & 0.013 & R\rlap{V} & nonmember & & \\
187 & F2\,V     & 0.424 & 0.038 & 2.75 & 7.68 & $-$0.06\rlap{:} & 0.008 & R\rlap{V} & SB2~ORB & & \\
189 & F3\,V     & 0.439 & 0.036 & 2.86 & 8.41 & $-$0.09\rlap{:} & 0.008 & & & & \\[5pt]
192 & F2\,IV    & 0.412 & 0.029 & 2.51 & 8.23 & $-${\em 0.06} & 0.008 & R\rlap{V} & & RO\rlap{T} & \\
193 & F1\,IV    & 0.403 & 0.035 & 2.15 & 8.05 & +{\em 0.07} & 0.010 & & & & \rlap{$\ast$} \\
196 & F2.5\,IV  & 0.432 & 0.039 & 2.13 & 8.12 & $-${\em 0.14} & 0.011 & & & & \\
197 & F5\,IV--V & 0.445 & 0.026 & 3.19 & 8.41 & $-$0.14 & 0.009 & & & & \\
205 & F3\,IV    & 0.432 & 0.038 & 2.69 & 7.21 & $-${\em 0.16} & 0.009 & R\rlap{V} & SB2~ORB & & \rlap{$\ast$} \\[5pt]
206 & F4\,IV    & 0.459 & 0.059 & 2.05 & 7.98 & $-${\em 0.05} & 0.010 & R\rlap{V} & SB1 & RO\rlap{T} & \\
209 & A0\,V     & 0.185 & 0.029 & 1.32 & 8.41 &                  & 0.011 & R\rlap{V} & SB & & \rlap{$\ast$} \\
213 & G9\,III   & 0.745 & 0.024 & 0.66 & 8.38 & $-$0.10 & 0.007 & R\rlap{V} & SB & & \\
214 & F1\,IV    & 0.398 & 0.042 & 2.25 & 8.21 & $+${\em 0.03} & 0.008 & R\rlap{V} & & RO\rlap{T} & \rlap{$\ast$} \\
217 & F2\,IV    & 0.431 & 0.038 & 2.17 & 8.26 & $+${\em 0.03} & 0.010 & & & & \\[5pt]
218 & F4\,IV    & 0.443 & 0.020 & 1.71 & 8.36 & $-${\em 0.06} & 0.009 & & & & \\
219 & F2\,IV--V & 0.413 & 0.024 & 2.65 & 7.85 & $-$0.06\rlap{:} & 0.008 & R\rlap{V} & SB2~ECL & & \rlap{$\ast$} \\
220 & K0\,IV    & 0.740 & 0.007 & 3.51 & 6.10 & $-$0.06 & 0.020 & & nonmember & & \\
222 & F2\,V     & 0.405 & 0.029 & 2.76 & 8.21 & $-${\em 0.13} & 0.008 & R\rlap{V} & & RO\rlap{T} & \\
226 & K1\,V     & 0.680 & 0.025 & 6.09 & 5.62 & $-$0.51 & 0.013 & & nonmember & & \\
\tablerule
\end{tabular}
}
\end{center}

\begin{center}
\vbox{\footnotesize\tabcolsep=5pt
\begin{tabular}{rlrrrrrrllll}
\multicolumn{12}{c}{\parbox{120mm}{\baselineskip=12pt {\normbf\ \
Table~2.}{\norm\ (continued)}}}\\
\tablerule \\[-6pt]
No. & Sp & \llap{($Y$$-$$V$)}$_0$ & $E$$_{Y-V}$ & $M_V\,$ & \llap{($m$}{\kern-0.05em}--{\kern-0.05em}$M$\rlap{)$_V$} & ~[Fe/H\rlap{]} & $\Delta Q$~ & \multispan{4}~Notes: RV{\kern-0.07em}, SB{\kern-0.07em}, ROT{\kern-0.07em}, etc\rlap{.} \\
\tablerule \\[-6pt]
232 & F3\,V     & 0.442 & 0.022 & 3.25 & 8.40 & $-$0.13 & 0.006 & & & & \\
234 & F3\,V     & 0.428 & 0.033 & 2.81 & 7.87 & {\em +0.03} & 0.009 & R\rlap{V} & & RO\rlap{T} & \rlap{$\ast$} \\
235 & F5\,V     & 0.470 & 0.024 & 3.38 & 8.02 & $-$0.20 & 0.008 & R\rlap{V} & SB2~ECL & & \rlap{$\ast$} \\
237 & F9\,V\,bin. & 0.524 & 0.036 & 3.72 & 8.61 & $-$0.16 & 0.010 & R\rlap{V} & SB1~ORB & & \\
238 & F4\,IV    & 0.447 & 0.044 & 2.32 & 7.65 & $-${\em 0.15} & 0.008 & R\rlap{V} & SB1~ORB & & \\[5pt]
254 & F1\,IV--V & 0.401 & 0.039 & 2.80 & 8.11 & $-${\em 0.13} & 0.008 & R\rlap{V} & SB1~ORB & RO\rlap{T} & \\
259 & F4\,V     & 0.444 & 0.028 & 3.04 & 8.34 & $-$0.13 & 0.012 & R\rlap{V} & & & \\
261 & F5\,V     & 0.459 & 0.023 & 3.42 & 7.75 & $-$0.08 & 0.006 & R\rlap{V} & & & \\
263 & F1\,IV    & 0.386 & 0.023 & 2.63 & 8.32 & $-${\em 0.14} & 0.011 & & & & \\
266 & F4\,IV--V & 0.434 & 0.022 & 2.84 & 8.37 & $-$0.17 & 0.007 & R\rlap{V} & SB1 & & \\[5pt]
295 & G8\,III   & 0.721 & 0.015 & 0.88 & 8.43 & $-$0.09 & 0.009 & R\rlap{V} & & & \\
300 & F2\,IV    & 0.425 & 0.029 & 2.31 & 7.29 & $-${\em 0.09} & 0.011 & R\rlap{V} & SB2~ORB & &\\
302 & F3\,V     & 0.420 & 0.017 & 3.13 & 8.39 & $-$0.12 & 0.008 & & & & \\
304 & F5\,V     & 0.456 & 0.028 & 3.41 & 8.46 & $-$0.06 & 0.010 & R\rlap{V} & SB1 & & \\
311 & G9\,III   & 0.765 & 0.038 & 0.67 & 8.40 & $-$0.09 & 0.007 & R\rlap{V} & & & \\[8pt]
\multispan{6}~~Additional stars with photometry from DP:~~~~~ & & & & & & \\[3pt]
  3 & G9\,III   & 0.76~  & 0.04~ & 1.06 & 8.48 & $-$0.29 & 0.019 & R\rlap{V} & & &  \rlap{$\ast$} \\
 24 & G9\,III   & 0.75~  & 0.02~ & 0.77 & 8.16 & $-$0.07 & 0.008 & R\rlap{V} & & & \\
 27 & G9\,III   & 0.75~  & 0.04~ & 0.78 & 8.36 & $-$0.23 & 0.010 & R\rlap{V} & & & \\
 48 & F7\,V     & 0.50~  & 0.03~ & 3.89 & 8.41 & $-$0.20 & 0.007 & R\rlap{V} & & & \\
 80 & F8\,V     & 0.52~  & 0.04~ & 3.78 & 9.01 & $-$0.09 & 0.012 & R\rlap{V} & & & \\[5pt]
176 & F6\,V     & 0.47~  & 0.02~ & 3.72 & 8.31 & $-$0.15 & 0.009 & R\rlap{V} & & & \\
184 & F8\,V     & 0.47~  & 0.00~ & 4.36 & 8.27 & +0.12\rlap{:} & 0.008 & R\rlap{V} & & & \\
208 & K0\,III   & 0.79~  & 0.02~ & 0.94 & 8.00 & $-$0.21 & 0.010 & R\rlap{V} & SB1~ORB & & \rlap{$\ast$} \\
293 & F5\,V     & 0.48~  & 0.03~ & 3.57 & 8.35 & $-$0.13 & 0.016 & R\rlap{V} & & & \\
P~\,\,937 & F2\,V & 0.42~ & 0.03~ & 2.59 & 8.37 & $-${\em 0.11} & 0.008 & & & & \\
P\,1263 & G9\,III & 0.75~ & 0.02~ & 0.71 & 8.28 & $-$0.07 & 0.012 & R\rlap{V} & & & \\
\tablerule
\end{tabular}

\begin{list}{}{}
\item[{\bf Notes:}]
\item[H\,3.]Detection of Li in the spectrum shows this star to be a first-ascent giant (van den Berg \& Verbunt 2001).
\item[H\,58.]Metallic-line star (Garrison 1972).
\item[H\,77.]Detection of Li in the spectrum shows this star to be a first-ascent giant (Pilachowski et al. 1988; Gilroy 1989).
\item[H\,193.]Metallic-line star (Garrison 1972).
\item[H\,205.]Optical identification for soft X-ray source (van den Berg \& Verbunt 2001).
\item[H\,208.]Detection of Li in the spectrum shows this star to be a first-ascent giant (Pilachowski et al. 1988).
\item[H\,209.]Blue straggler. Optical identification for soft X-ray source (van den Berg \& Verbunt 2001).
\item[H\,214.]Optical identification for soft X-ray source (van den Berg \& Verbunt 2001).
\item[H\,219.]The eclipsing binary DS\,And ($P$=1.01$^{\rm d}$); for more details, see Schiller \& Milone (1988).
\item[H\,234.]Metallic-line star (Garrison 1972).
\item[H\,235.]The eclipsing binary, identified as a contact system with a variable period ($P$\,$\approx$\,0.41$^{\rm d}$); for more details, see Milone et al.
(1995) and Milone \& Terrell (1996). Optical identification for soft
X-ray source (van den Berg \& Verbunt 2001).
\end{list}
}
\end{center}

\noindent reddening $E_{Y-V}$\,=\,0.027 found for the cluster
(\S\,4.1). The [Fe/H] values are averages over the estimates from
$P$$-$$X$, $X$$-$$Y$ and $P$$-$$Y$ calibrations, but, in the cases
where the calibrations do not apply we give instead the estimates
extracted from ``photometric boxes" (these values are given in
italics). In the eighth column, labeled ``$\Delta Q$", we give the
average $\Delta Q$ (Eq.~2) in the ``photometric box" as indicative
of the accuracy of the parameters determined by this method. The
last columns of the table contain brief notes from the literature:
RV indicates that the star has repeated radial-velocity
observations, SB1/SB2/ORB/ECL indicates its binarity status, ROT
means rapid rotation (i.e., projected rotational velocities over 30
km\,s$^{-1}$), and an asterisk in the final column refers to
additional comments at the end of the table. The lower 11 lines of
Table~2 present the photometric parameters for 11 additional stars,
five red giants and six main-sequence F stars, which were measured
in the {\em Vilnius} system by DP. We found in DP a total of 17
members of the cluster, which were not observed by us, but only 11
of these could be classified by the ``photometric box" method with
sufficient accuracy ($\Delta Q$ or $\Delta (CI)_0$ $\leq$\,0.02).
The numbers of the 11 stars are also of Heinemann, except for the
two last stars marked with ``P", for which the numbering system of
Platais (1991) is adopted.

The stars H\,186, H\,220 and H\,226, known as proper-motion and/or
radial-velocity nonmembers (Pilachowski et al. 1988; Platais 1991;
DLMT), will be excluded from further consideration. According to our
photometric classification, all of the three stars lie in the
foreground of the cluster: at distances 210, 160 and 130 pc,
respectively.

Known spectroscopic binaries, 18 among the main-sequence stars and
four among red giants, constitute a fraction of 30\% of the sample
of 73 member stars in Table~2. However, not all of the main-sequence
stars have radial velocity observations and hence not checked for
binarity. If to consider only those 34 F-type stars which have
multiple radial velocity measures, the percentage of the
spectroscopic binaries among them is raised to 53\%. We have to keep
in mind that photometric estimates of [Fe/H] and distance moduli for
binaries and, at least, part of the stars with unknown radial
velocities can be affected by binarity.

Another serious problem is that rapid axial rotation, which might be
expected in stars near the main-sequence turnoff (earlier than
spectral type F5), can affect their luminosities and colors (see,
e.g., Roxburgh \& Strittmatter 1965). In Table~2, we have 11 F-type
stars with projected rotational velocities found to be over 30
km\,s$^{-1}$ (Gunn \& Kraft 1963) or whose radial velocities were
derived by DLMT using computed templates with rotational velocities
greater than 30 km\,s$^{-1}$. The photometric effects of rapid
rotation were noticeable in deriving metallicities. The mean [Fe/H]
for the 11 rapid rotators is found to be $-$0.05\,$\pm$\,0.07(s.d.),
which is higher than that for stars supposedly unaffected by
binarity and rotation effects (i.e., lying along the lower envelope
of the main sequence): $-$0.14\,$\pm$\,0.03 and
$-$0.14\,$\pm$\,0.10, considering [Fe/H] determined through the
calibrations and from ``boxes", respectively. As it has been shown
by {\v Z}itkevi{\v c}ius \& Strai{\v z}ys (1972), the effect of
axial rotation is most noticeable on the color indices $U$$-$$P$,
$P$$-$$X$ and $P$$-$$Y$ and not so on $X$$-$$Y$. With increasing
rotation, both $P$$-$$X$ and $P$$-$$Y$ get larger, and we would
therefore expect to arrive at higher values of photometric
metallicity.

Perhaps the most visible manifestation of binary stars and rapid
rotators present in the cluster is the width of the observed
main sequence (see Figure~2).

\begin{figure}[t]
\centerline{\psfig{figure=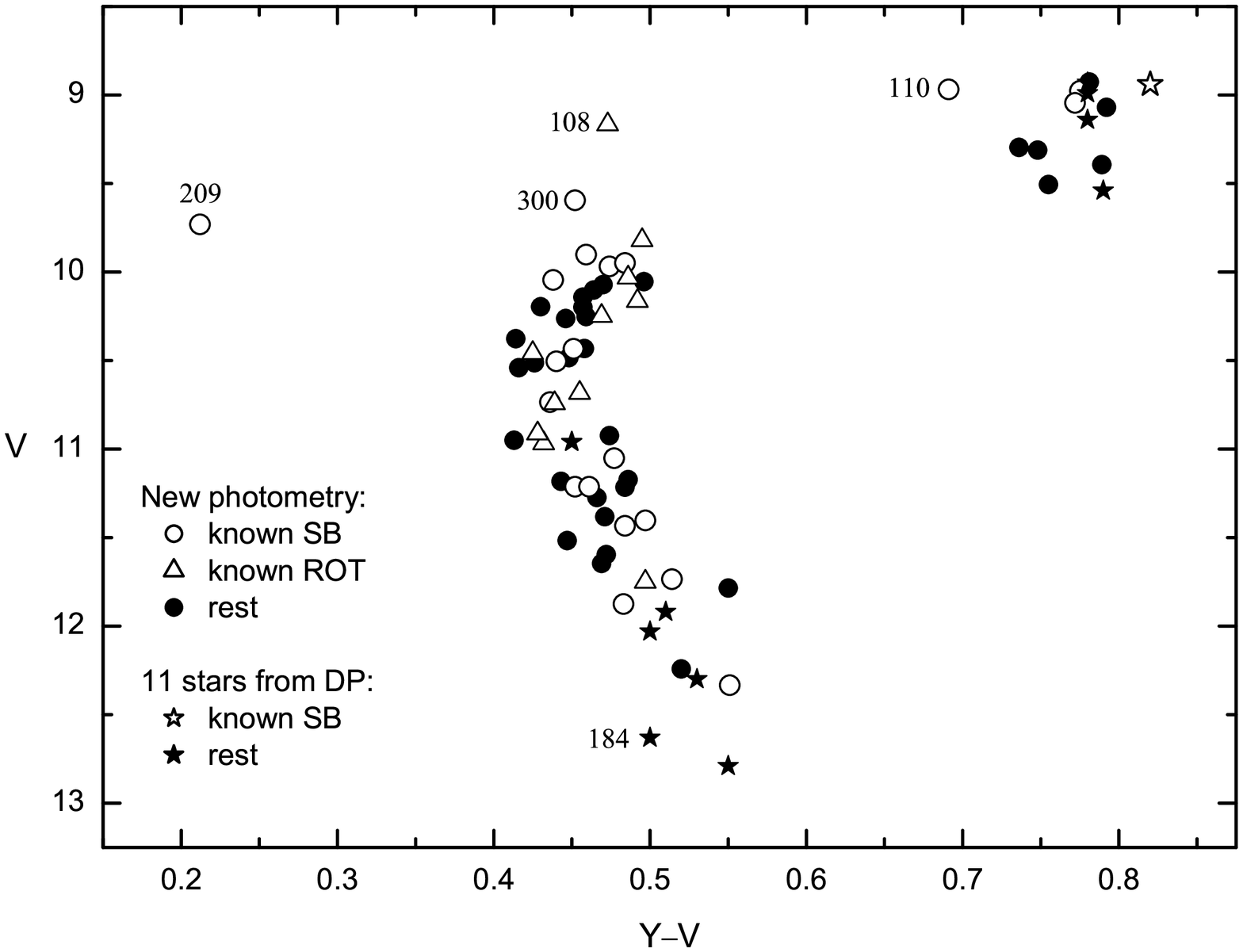,width=115truemm,angle=0,clip=}}
\vskip2mm \captionb{2}{CMD for the member stars observed in
NGC\,752. An additional 11 stars are taken from DP, with these
points denoted by star-like symbols. The blue straggler (H\,209) and
a few most deviating stars are marked with their numbers.}
\end{figure}

\noindent Therefore, in deriving the main
cluster parameters from individual stars (next section), we have
taken into account whether or not any estimate is based on
single-star candidates and stars with small rotational velocities.

\sectionb{4}{RESULTS AND DISCUSSION}

\subsectionb{4.1}{Cluster reddening and distance}

To obtain the reddening and the distance appropriate to the cluster,
we averaged individual color excesses $E_{Y-V}$ and apparent
distance moduli $(m$$-$$M)_V$ from Table~2 separately for the RGB
stars and the main sequence (F-type) stars. In an effort to reduce,
or at least discern, the influence of photometric effects of
binarity and rotation on the average values to be determined, the
following groups of stars have been considered: red giants with
spectroscopic binaries excluded ({\em g1}) and included ({\em g2}),
F stars which define the lower envelope of the observed main
sequence ({\em f1}), and F stars with nonvariable radial velocities
and small or moderate rotation ({\em f2}). The two faintest stars
from DP, H\,80 and H\,184, were not included in the `envelope' group
{\em f1} because of large differences in $V$ magnitude between our
data and those of DP at the fainter magnitude end. The groups of
F stars known as spectroscopic binaries ({\em f3}) and rapid
rotators ({\em f4}) were taken into consideration for comparison
purposes only.

The resulting mean values, together with their errors, are given in
Table~3. We\break
\\[-29pt]
\begin{center}
\vbox{\footnotesize\tabcolsep=4pt
\begin{tabular}{clrclcl}
\multicolumn{7}{c}{\parbox{115mm}{\baselineskip=9pt {\smallbf\ \
Table~3.}{\small\ Mean values of reddening and apparent distance
moduli for different groups of stars in
NGC\,752.}}}\\[6pt]
\tablerule\\[-8pt]
& Groups of stars & \multicolumn{1}{c}{$N$} &
\multicolumn{1}{c}{$E_{Y-V}$} & \multicolumn{1}{c}{s.d.} &
\multicolumn{1}{c}{($m$$-$$M$)$_V$} & \multicolumn{1}{c}{s.d.}
\\[-1pt]
\tablerule\\[-8pt]
{\itl g1} & Red giants: binaries (SB) excluded & 10 & 0.025\,$\pm$\,0.003 & 0.011 & 8.40\,$\pm$\,0.05 & 0.15 \\
{\itl g2} & Red giants: all               & 14 & 0.024\,$\pm$\,0.003 & 0.010~~ & 8.36\,$\pm$\,0.05 & 0.17 \\[5pt]
{\itl f1} & F stars: the lower envelope of MS & 16 & 0.028\,$\pm$\,0.002 & 0.009 & 8.37\,$\pm$\,0.04 & 0.14 \\
{\itl f2} & F stars with RV const: ROT exclude\rlap{d} & 9 & 0.025\,$\pm$\,0.004 & 0.011 & 8.34\,$\pm$\,0.13 & 0.40 \\[5pt]
{\itl f3} & F stars: known binaries (SB1,\,SB2)& 18 & 0.033\,$\pm$\,0.003 & 0.011 & 7.98\,$\pm$\,0.10 & 0.42 \\
{\itl f4} & F stars: rapid rotators (ROT) & 11 & 0.039\,$\pm$\,0.003 & 0.009 & 8.06\,$\pm$\,0.09 & 0.29\\[5pt]
& \multispan{2}{\itl ~Parameters assumed for the cluster}:~~~ &&&&\\[1pt]
{\itl g1}+{\itl f1} & Red giants and F stars & 26 & 0.027\,$\pm$\,0.002 & 0.010 & 8.38\,$\pm$\,0.03 & 0.14 \\
 \tablerule
\end{tabular}
\begin{list}{}{}
\item[] The error following the mean refers to the standard error of mean (s.e.m.) and s.d. is the standard deviation for one star.
\end{list}
}
\end{center}

\noindent note that standard deviations from the mean for one star
(s.d.) can be considered here as an independent test of the internal
accuracy of our estimates from {\em Vilnius} photometry of the color
excess $E_{Y-V}$ and $M_V$.

As we can see from Table~3, the average values of $E_{Y-V}$ and
distance moduli for the G--K giants (groups {\em g1} and {\em g2})
are close to the corresponding mean values for the F-type stars
which are supposed to be unaffected by binarity and rotation effects
(groups {\em f1} and {\em f2}). As one would expect, the groups of
spectroscopic binaries ({\em f3}) and rapid rotators ({\em f4}) show
a systematic difference in both the mean $E_{Y-V}$ and
$(m$$-$$M)_V$. Since the presence of the companion makes the system
brighter and the change in gravity sensitive color indices leads to
a somewhat fainter luminosity, the distance moduli for binaries
become smaller than those derived for single stars. Rapid stellar
rotation mimics photometrically the effect of binarity (see, e.g.
Fig.~2 in Maeder 1974), leading also to smaller distance moduli.

\begin{wrapfigure}[18]{r}[0pt]{63mm}
\vskip1mm\psfig{figure=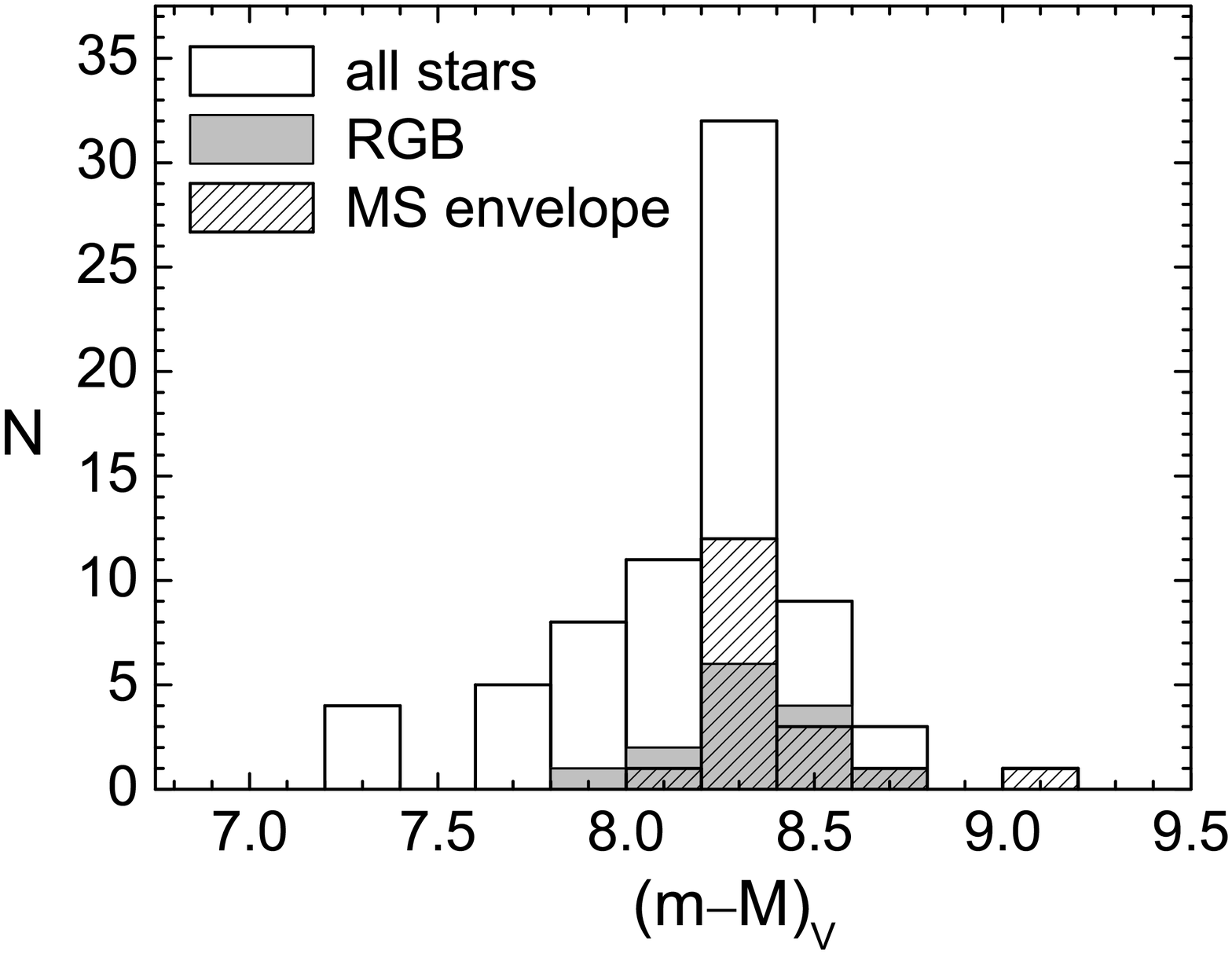,width=61truemm,angle=0,clip=}
\vskip1mm \captionb{3}{The distribution of individual distance
moduli.}
\end{wrapfigure}
The red giants ({\em g1} and {\em g2}) and the F-type stars lying on
the lower boundary of the main sequence ({\em f1}) give
self-consistent values of the apparent distance modulus (see, also,
Figure~3). For the F-star group {\em f\,2}, the mean distance
modulus is similar, but the standard deviation from the mean, around
$\pm$0.4 mag, is much larger than for the above groups of stars.
Therefore, as appropriate values for the cluster reddening and
distance modulus we assumed the average over the combined subsample
of the single red giant stars ({\em g1}) and F stars along the lower
envelope of the main sequence ({\em f1}). From an average over 26
such stars, we have the mean foreground reddening
$E_{Y-V}$=\,0.027\,$\pm$\,0.002\,(s.e.m.), with a standard deviation
of 0.010 mag for one star. Taking the $E_{Y-V}/E_{B-V}$ ratios as a
function of spectral type from Table~65 of Strai{\v z}ys (1992) and
the ratio $R_{YV}$\,=\,$A_V/E_{Y-V}$\,=\,4.16 (Kazlauskas 1996), we
arrive at the mean value $E_{B-V}$=\,0.034\,$\pm$\,0.013\,(s.d.) and
the total extinction $A_V$\,=\,0.11 mag. The mean apparent distance
modulus, $(m$$-$$M)_V$, based on the same 26 stars, is
8.38$\pm$\,0.03 (s.e.m.), with a standard deviation of 0.014 mag
for one star. With the above value of $A_V$ we obtain the true
distance modulus $(m$$-$$M)_0$=\,8.27$\pm\,$0.03 (s.e.m.). This
leads to the cluster distance 450\,$\pm$\,10 pc. However, the
theoretical isochrone fitting to the CMD, to be described in
\S\,4.3.2, gave the distance modulus by 0.2 mag smaller than the
above value.

\vskip 20pt

\begin{center}
\vbox{\footnotesize\tabcolsep=5pt
\begin{tabular}{llll}
\multicolumn{4}{c}{\parbox{100mm}{\baselineskip=9pt {\smallbf\ \
Table~4.}{\small\ Determinations of interstellar reddening and true
distance modulus for NGC\,752.}}}\\[6pt]
\tablerule\\[-9pt]
$E_{B-V}$ & $(m-M)_0$ & Method  & Reference  \\
\tablerule
 0.03  &  & $B$--$V$,\,MK-type relation & Johnson (1953) \\
 0.024 & 8.2\,~$\pm$0.5 & $B$--$V$,\,MK-type relation & Roman (1955) \\
 0.025$\pm$0.009 & & {\it UBV} & Gunn \& Kraft (1963) \\
 0.04  & 7.80 & {\it UBV} & Eggen \& Sandage (1964) \\
 0.04  & 8.2$\pm$0.1$^\dagger$ & {\it UBV},~{\it BV}  & Cannon (1970) \\
 0.036$\pm$0.014 & 8.0\,~$\pm$0.3 & {\it uvby}$\beta$ & Crawford \& Barnes (1970) \\
 0.005$\pm$0.025 & 7.8\,~$\pm$0.3 & {\it uvby}$\beta$ & Bell (1972) \\
 0.015 & & $B$--$V$,\,MK-type relation & Garrison (1972) \\
 0.04  & 7.7\,~$\pm$0.1 & {\it UBViyz} & Jennens \& Helfer (1975) \\
 0.016$\pm$0.008 & 7.9\,~$\pm$0.2 & {\it DDO} & Hardy (1979) \\
 0.046$\pm$0.019 & 7.83$\pm$0.28 & {\it Geneva} & Nicolet (1981b) \\
 0.04\,~$\pm$0.01 & 8.2\,~$\pm$0.2 & {\it uvby}$\beta, {\it DDO}$ & Twarog (1983) \\
 0.036$\pm$0.014 & 8.12 & {\it uvby}$\beta$ & Nissen (1988) \\
 0.04\,~$\pm$0.03 & & $T_{\rm eff}$,\,($B$-$V$)$_0$ relation & Hobbs \& Thorburn
 (1992) \\
 0.025$\pm$0.025 & 8.1$^\dagger$ & {\it Vilnius} & Dzervitis \& Paupers (1993) (DP) \\
 0.035$\pm$0.005$^*$ & 8.25$\pm$0.10$^*$ & combined photometry & Daniel et al. (1994) (DLMT) \\
 & 8.08$\pm$0.08$^*$ & eclips. binaries: RV, {\it BVR}\rlap{$I$} & Milone \& Terrell (1996) \\
 0.04 & 8.22 & {\it DDO}, merged sources & Twarog et al. (1997) \\[5pt]
 0.034$\pm$0.013 & 8.27$\pm$0.14 & {\it Vilnius}: individual stars & present paper \\
 0.034 & 8.06 & {\it Vilnius}: isochrone fitting & present paper (\S\,4.3.2) \\
\tablerule
\end{tabular}
\begin{list}{}{}
\item[{\bf Notes.}]The errors quoted are the standard deviations from the mean for one star,
except for those marked with an asterisk, which are probable errors
of the weighted mean.

The distance moduli marked with a dagger are given assuming the
modulus for the Hyades 3.3 mag.
\end{list}
}
\end{center}

Most of the available literature estimates of $E_{B-V}$ and
$(m$$-$$M)_0$ for NGC\,752 are listed in Table~4. It is apparent
that there is satisfactory agreement among the many determinations
of the interstellar reddening. We note that our mean value of
$E_{B-V}$ is slightly larger than that obtained earlier from {\em
Vilnius} observations by DP, but still consistent within the errors.
Our result is in good agreement with that by DLMT,
$E_{B-V}$=\,0.035$\pm$\,0.005\,(p.e.), based on the combined data in
several photometric systems.

While most of the reddening values are quite consistent, the
estimates of the distance modulus, at least those summarized in
Table~4, range from 7.7 to 8.3 mag. Our estimate of $(m$$-$$M)_0$
from individual stars appears to be the closest to the result by
DLMT. From earlier {\em Vilnius} photometry in NGC\,752, DP obtained
the true distance modulus 8.1$\pm$\,0.15, which is in much better
agreement with our value to be derived in \S\,4.3.2 by using an
elaborate fitting procedure. Milone \& Terrell (1996) from {\it BVRI}
photometry and radial velocity curves of the two eclipsing binaries,
H\,219 and H\,235, have also obtained the mean distance modulus
somewhat smaller than given in recent photometric works:
$(m$$-$$M)_0$=\,8.08\,$\pm$\,0.08 (with $E_{B-V}$=\,0.033).

\vskip2mm

\subsectionb{4.2}{Metallicity}

We shall consider only those cluster stars which have metallicities
derived through calibrations of color indices (i.e., with [Fe/H]
values in Table~2 given {\em not} in italics). The 33 such stars,
when treated together, have a mean of
[Fe/H]\,=\,$-$0.12\,$\pm$\break 0.07 (s.d.). The standard deviation
for one star is rather small, therefore it is likely that our
estimates of [Fe/H] can be accurate enough to be used to discover
possible evolutionary differences in metallicity, as it was done in
the paper by Barta\v si\= ut\. e \& Tautvai{\v s}ien{\. e} (2004)
for the the intermediate-age cluster NGC\,7789. However, NGC\,752
does not display an extended red-giant branch, and we are not able,
like in the case of NGC\,7789, to compare photometric metallicities
of the clump giants with those of the less evolved first-ascent RGB
stars. Instead, we can attempt a comparison of [Fe/H] between the
evolved and unevolved stars represented by the red giants of the
clump and the main-sequence stars, respectively.

The red giants in NGC\,752 are distributed almost uniformly in the
region between $M_V$\,=\,$+$0.6 and $+$1.1 mag, therefore it is
difficult to distinguish by their location in the CMD (see Figure~2)
which of the giants are on their first-ascent of the RGB and which
are in the clump. The one plausible criterion for identifying the
evolutionary status of red giants is Li abundance in their
atmospheres: the lithium left at the end of the main sequence
evolution would be still present in first-ascent giants but must
have been destroyed after evolving past the first-ascent branch
(Pilachowski et al. 1988). Of the 14 red giants included in Table~2,
only three stars are known to have lithium detected (Pilachowski et
al. 1988; Gilroy 1989; van~den~Berg \& Verbunt 2001): H\,3, H\,77
and H\,208, and these can be on the first-ascent branch. The
evolutionary status of the single-lined binary H\,110, which is
found in an unusual location in the CMD and had been coined ``red
straggler" by Eggen \& Iben (1988), is unclear. According to
Mermilliod et al. (1988) this binary could consist of the red giant
primary, which might have evolved past the giant tip and underwent a
significant mass loss (and/or mass transfer), and a dwarf secondary.
Excluding the ``red straggler" and the above three first-ascent
giants, the clump is represented by the remaining 10 red giant
stars, with the average absolute magnitude, according to Table~2,
$M_V$=$+$0.73\,$\pm$\,0.09 (s.d.).

The 10 clump red giant stars have a mean of
[Fe/H]\,=\,$-$0.08\,$\pm$\,0.03 (s.e.m.) with a standard deviation of
0.08 dex, independent of whether the two known spectroscopic
binaries are included among them or not.

The average for the 21 F-type stars on the main sequence is found to
be [Fe/H]\,=\,$-$0.13\,$\pm$\,0.01 (s.e.m.), with a standard
deviation of 0.04 dex. Excluding the known spectroscopic binaries,
the remaining 13 main-sequence stars have the same mean,
$-$0.13\,$\pm$\,0.03 (s.d.). The eight stars representing the
single-star domain in the CMD, i.e., the lower envelope of the
observed main sequence, have a mean of
[Fe/H]\,=\,$-$0.14\,$\pm$\,0.01 (s.e.m.), with a standard deviation
of 0.03 dex.  Considering F-type stars with radial velocities not
reported as variable, we arrive again at the same mean:
[Fe/H]\,=\,$-$0.14\,$\pm$\,0.04 (s.d.). In either of the above cases,
the main-sequence stars exhibit a slightly lower metallicity than
the red clump giants.

Although this difference in [Fe/H] is small (significant formally at
the 2$\sigma$ level), nevertheless it reminds the one found by
Barta\v si\= ut\. e \& Tautvai{\v s}ien{\. e} (2004) between the
first-ascent red giants and the clump stars in the intermediate age
cluster NGC\,7789. The interpretation given in their paper was that
the difference might be an indication of extra-mixing on the RGB
ascent. Such a mixing process could have caused additional
CN-processed elements to be transported to the surface, hence
inducing a change in the surface composition (see, e.g., Gilroy
1989; Gilroy \& Brown 1991). In the case of red giant stars, the
{\em Vilnius} passbands $P$ (360--390 nm), $X$ (390--420 nm) and $Y$
(450--480 nm), used for the determination of [Fe/H], contain a
number of CN, CH and C$_2$ molecular bands the spectral features of
which can be affected by an extra-mixing process (see Tautvai{\v
s}ien{\. e} et al. 2003). If strong enough, such spectral changes
can influence photometric metallicity determinations. Therefore, the
chemical abundance of the red clump giants need not represent the
zero-age surface abundance of the cluster. In NGC\,752, however,
there are exceedingly few first-ascent giant candidates, so we are
unable to extract a reliable mean value of photometric metallicity
which could be compared with that found for the clump. The only
three first-ascent giant candidates in Table~2 (H\,3, H\,77 and H\,208)
have an average of [Fe/H]\,=\,$-$0.20, which is lower than that for the
clump giants, thus confirming the above interpretation. We note,
however, that for H\,3 the color indices taken from DP may be of
lower accuracy ($\Delta$$Q$\,=\,0.019, i.e., twice as much as for any
other red giant star in Table~2), and thus the low [Fe/H] value
derived for this red giant, $-$0.29 dex, needs to be treated with
reserve. The remaining two first ascent giants give an average
[Fe/H] of $-$0.15 dex, still close to that for the main-sequence
stars which are thought to have undergone little or no mixing.

A summary of [Fe/H] determinations available in the literature is
presented in Table~5, together with our values obtained separately
for the red clump giants and the F-type stars lying near the lower
envelope of the main sequence. As can be seen from the table, the
estimates based on different studies vary approximately from solar
metallicity to $-$0.2 dex. A thorough reanalysis of the cluster
metallicity was made by DLMT who arrived at a weighted mean of
[Fe/H]\,=\,$-$0.15\break $\pm$\,0.05 (p.e.) from both photometric and
spectroscopic approaches. Our estimate for the unevolved cluster
stars is in good agreement with this value. However, the most recent
studies on metal abundance of the main-sequence stars give [Fe/H]
values much closer to the solar metallicity. Sestito et al. (2004),
for example, from high-dispersion spectra of 18 main-sequence G-type
stars obtained [Fe/H]\,=$+$0.01\,$\pm$\,0.04, based on comparison
with the Hyades for which [Fe/H]\,=\break $+$0.12 was adopted. With
the same value set for the Hyades, Anthony-Twarog \& Twarog (2006)
arrived at a slightly subsolar value,
[Fe/H]\,=\,$-$0.06\,$\pm$\,0.09, from the new {\it uvbyCa} photometry of
10 F-type stars. To check whether or not our [Fe/H] values have the
same zero point as that set in the above two works, we took 21
members of the Hyades cluster with constant radial velocities, which
were observed in the {\em Vilnius} system by Dz{\`e}rv{\`i}tis \&
Paupers (1994), and derived metallicities using exactly the same
calibrations as for NGC\,752. An average over these 21 Hyades dwarfs
was found to be of [Fe/H]=$+$0.13\,$\pm$\,0.10\,(s.d.), nearly the
same as that adopted in the above two papers. Therefore, according
to {\em Vilnius} photometry, NGC\,752 is likely to have the
metallicity by about 0.25 dex lower than that of the Hyades. We will
use in further analysis [Fe/H]\,=\,$-$0.14$\pm$\,0.03 (s.d.), an
average of estimates for F-type stars on the lower envelope of the
main sequence, as the metallicity for the cluster as a whole.

\begin{center}
\vbox{\footnotesize\tabcolsep=4pt
\begin{tabular}{llrll}
\multicolumn{5}{c}{\parbox{92mm}{\baselineskip=9pt {\smallbf\ \
Table~5.}{\small\ Determinations of metallicity for
the cluster NGC\,752.}}}\\[3pt]
\tablerule\\[-6pt]
[Fe/H]$^\dagger$ & Stars & $N$ & Method  & Reference  \\
\tablerule
 $-$0.2$^{\rm a}$  & & & $UBV$  & Arp (1962) \\
 $-$0.2 & & 45 & $UBV$ & Eggen (1963) \\
 $+$0.02$\pm$0.15 & F-type & 2 & High dispersion & Gunn \& Kraft (1963) \\
 $-$0.2 & RG~(H\,213) & 1 & High dispersion & Wallerstein \& Conti (1964) \\
 $+$0.2$\pm$0.2 & F-type & 31 &$uvby\beta$ & Bell (1972) \\
 $-$0.2 & & 16 &$UBViyz$ & Jennens \& Helfer (1975) \\
 $-$0.22$^{\rm a{\kern-0.05em},b}$ & & & $UBV$,\,$DDO$ & Hirshfeld et al. (1978) \\
 $-$0.1$\pm$0.1 & RG& 11 &$DDO$ & Hardy (1979) \\
 $+$0.1$\pm0.25$ & RG~(H\,213) & 1 &High dispersion & Komarov \& Shcherback (198\rlap{0)} \\
 $-$0.21$\pm$0.09$^{\rm a{\kern-0.05em},b}$ & RG, F-type & 35 &$uvby\beta$,\,$DDO$,\,$UBV$ & Twarog (1983) \\
 $-$0.15$\pm$0.2 & RG~(H\,213) & 1 & High dispersion & Geisler (1984) \\
 $-$0.15 & RG & 4 & $ Washington$ & Canterna et al. (1986) \\
 $-$0.05$\pm$0.13 & F-type & 26 &$uvby\beta$ & Nissen (1988) \\
 $-$0.09$\pm$0.05 & F-type &  8 & High dispersion & Hobbs \& Thorburn
(1992) \\
 $-$0.16$\pm$0.05 & RG & 9 & Moderate dispersion & Friel \& Janes (1993) \\
 $-$0.15$\pm$0.05$^{\rm a{\kern-0.05em},c}$ &  & & Combined data & Daniel et al. 1994) (DLMT) \\
 $+$0.01$\pm$0.04$^{\rm b}$ & MS G-type & 18 & High dispersion & Sestito et al. (2004) \\
 $-$0.06$\pm$0.09$^{\rm b}$ & F-type & 10 & $uvbyCa$ & Anthony-Twarog \& Twarog (2006)\\[5pt]
 $-$0.08$\pm$0.08 & RG clump & 10 & $Vilnius$ & present paper \\ 
 $-$0.14$\pm$0.03 & F-type & 8 & $Vilnius$ & present paper \\
\tablerule
\end{tabular}
\begin{list}{}{}
\item[$^\dagger$]The error quoted is the standard deviation from the mean for one star unless otherwise noted, or, when only one or two stars measured, the maximum uncertainty in [Fe/H] determination.
\item[$^{\rm a}$]From analysis of the available data.
\item[$^{\rm b}$][Fe/H] normalized to the Sun adopting for the Hyades
[Fe/H]\,=\,+0.12.
\item[$^{\rm c}$]The weighted mean and its probable error
from reanalysis of the available photometric and spectroscopic data,
adopting $E_{B-V}$\,=\,0.035.
\end{list}
}
\end{center}

\subsectionb{4.3}{CMD fitting and the binary population}

\subsubsectionb{4.3.1}{Description of the method and approach}

As is apparent from Figure~2, the main sequence of the cluster is
appreciably widened by the presence of unresolved binary stars and
rapid rotators. This causes a major complication in deriving the age
from fitting the CMD to the theoretical isochrones.

The photometric effect of binarity makes the star appear in the CMD
above the main sequence since the composite magnitude brightens by a
certain amount which depends on the mass ratio of the components. If
a binary is composed of two equal-mass components, it will have the
same color but lie at 0.75 mag above the main sequence. As one
lowers the mass ratio, the composite magnitude becomes closer to the
main sequence but the color of the system is redder than that of the
primary component alone. This effect can be used for the
identification of binary stars in the clusters, by applying, for
example, a method given by Romani \& Weinberg (1991).

In the case of multicolor photometry we can additionally use
different CMDs which provide different effects upon the colors. In
the {\em Vilnius} system, for example, the effects of the cooler
secondaries will affect the color in the $V$,\,$Y$$-$$S$ diagram
more than they will the color in the $V$,\,$Y$$-$$V$ diagram, i.e.,
the spread covering the binary trajectory reaches 0.1 mag along the
$(Y$$-$$S)$-axis and 0.05 mag along the $(Y$$-$$V)$-axis. The larger
the spread in the CMD, the easier becomes the identification of
unresolved binaries. We considered in this paper the photometric
effects upon all seven magnitudes of the {\em Vilnius} system. Note
that the red giant stars in our sample, all having a firmly
established binarity status from radial-velocity observations (see
Mermilliod et al. 1998), will not be considered here.

The method applied consisted of two steps: (1) producing artificial
(model) binaries for given metallicity, age and masses of the
components $({\mathrm{[Fe/H]}},\,t,\,m_1,\,m_2)$; (2) applying the
least-squares minimization to find for each observed star the model
binaries which provide the best match between the model magnitudes
and colors and those of the observed star.

The artificial binaries can be created by simple addition of the
fluxes of two stars having the same age, metallicity and all
possible mass ratios. The composite absolute magnitude $M_k$ of an
artificial binary is then
\begin{equation}
 M_k({\mathrm{[Fe/H]}},t,m_1,m_2) = -2.5\,\log\left(10^{-0.4\,M_k({\mathrm{[Fe/H]}},\,t,\,m_1)}+10^{-0.4\,M_k({\rm[Fe/H]},\,t,\,m_2)}\right)\,,
   \end{equation}
where $M_k({\mathrm{[Fe/H]}},t,m_1)$ is the absolute magnitude in
passband $k$ of a model primary of mass $m_1$, having age $t$ and
metallicity [Fe/H], and $M_k({\mathrm{[Fe/H]}},t,m_2)$ is the same
magnitude of a model secondary of mass $m_2$ and of the same age and
metallicity. In our case,
$k$\,=\,($U$,\,$P$,\,$X$,\,$Y$,\,$Z$,\,$V$,\,$S$) are the seven
filters of the {\em Vilnius} system. The composite colors of
artificial pairs are produced in a straightforward manner. Given the
values of the cluster foreground reddening and distance moduli, we
can instead produce the apparent magnitudes and colors.

Follow-up application of the least-squares minimization algorithm
allows us to find for each observed star the best match between its
magnitudes and colors and those of model pairs by varying the
assumed distance modulus, metallicity and age. Let the magnitudes in
the seven {\em Vilnius} filters of the $i$$^{\mathrm{th}}$ observed
star be ($u_i$,\,$p_i$,\,$x_i$,\,$y_i$,\,$z_i$,\,$v_i$,\,$s_i$) and
the seven magnitudes of a model binary star correspondingly
($U({\mathrm{[Fe/H]}},t,m_{1i},m_{2i}),\,.\,.\,.\,,\,S({\mathrm{[Fe/H]}},t,m_{1i},m_{2i})$).
Then, the least-squares minimization problem may be written as
\begin{equation}
\cases{
  \begin{array}{l}
\left(u_1\,\,-U({\mathrm{[Fe/H]}},t,m_{11},m_{21})\right)^2\,\,\,+...+\left(s_1\,\,-S({\mathrm{[Fe/H]}},t,m_{11},m_{21})\right)^2~\,=min
\\
.\,.\,.\\
\left(u_N-U({\mathrm{[Fe/H]}},t,m_{1N},m_{2N})\right)^2+...+\left(s_N-S({\mathrm{[Fe/H]}},t,m_{1N},m_{2N})\right)^2=min
  \end{array}
}
\end{equation}
which can be solved for a set of the model stars at various
distances by matching within the errors the magnitudes and colors of
the observed stars. Such a set found can be considered as analogous
to the observed sample and used in further analysis of the binary
population in the cluster.

To produce the artificial stars of a given [Fe/H], age and mass, we
used the Padova theoretical isochrones, computed in the
observational plane of the {\em Vilnius} system by Bressan \&
Tautvai{\v s}ien{\. e} (1996) on the basis of stellar evolutionary
models by Bressan et al. (1993) and Fagotto et al. (1994) with mild
overshoot from convective cores. However, for most of the colors in
the original tabulations by Bressan \& Tautvai{\v s}ien{\. e}, there
are appreciable systematic differences between the model data and
the fiducial lines derived empirically for the main sequence and RGB
stars in the {\em Vilnius} system. In a direct comparison of the
model colors at the age and composition of the Sun with the
corresponding {\em Vilnius} colors of the empirical fiducial lines
for solar-composition stars from Tables 66--68 of Strai{\v z}ys
(1992), the zero-point offsets in the color-temperature and
color-gravity relations have been determined and applied to the
original isochrones to provide a somewhat better match. The amount
of the zero-point shift ranged from $+$0.04 mag (the largest shift;
for $U$$-$$P$) to $+$0.01 mag (the smallest shift; for $Z$$-$$V$).

To have a set of theoretical isochrones with an extremely dense
coverage of metallicity and age appropriate to the cluster, the
Padova isochrones have been interpolated in 0.001 steps of $Z$
between $Z$\,=\,0.020 ($Y$=\,0.28) and $Z$\,=\,0.008 ($Y$=\,0.25)
and in 0.01 steps of $\log t$ between ages $\log t$\,=\,9.0, 9.1,
9.2 and 9.3. For the transformation of $Z$ to [Fe/H] and vice versa,
we assumed for the Sun $Z_\odot$\,=\,0.020, i.e., the same value as
in the Padova models.

The least-squares minimization procedure was performed by setting a
fixed value of reddening, corresponding to $E_{Y-V}$=\,0.027 derived
in \S\,4.1, and varying the other three parameters: distance modulus
(in 0.01 mag steps between $(m$$-$$M)_0$=\,8.0 and 8.4), metallicity
and age. Here, the width of the ``single-star" main sequence along a
theoretical isochrone was assumed to be consistent with the errors
in our photometry ($\pm$0.015 in the color-axis and $\pm$0.03 in the
magnitude-axis). Therefore, the cluster stars which appeared to fall
in a theoretical CMD within the boundaries of such a sequence were
treated as ``single". We remind that the minimization technique and
the CMD fit were applied only to the stars on the main sequence
($N$=\,58) and not on the RGB. The effects of axial rotation, which
could also affect the position of the star in the CMD, were
neglected in the present treatment.

It should be noted that the results of the least-squares
minimization described above depend critically on the theoretical
model adopted, and even more so on the choice of
temperature-to-color and gravity-to-color transformations. As it was
mentioned above, the {\em Vilnius} colors predicted in the original
isochrone tabulations of Bressan \& Tautvai{\v s}ien{\. e} (1996)
are not in satisfactory agreement with empirical data, thus being by
far the most important source of uncertainties in the photometric
identification of unresolved binaries and the best fit to the
observational data.

\vskip2mm

\subsubsectionb{4.3.2}{Best-fit results}

By varying the distance modulus, metallicity and age, the
least-squares minimization with a fixed value of reddening
($E_{Y-V}$=\,0.027) yields the following parameters for the cluster:
$(m$$-$$M)_V$=\,8.18\,$\pm$\,0.03, or a distance of 410 pc, the
metallicity $Z$\,=\,0.015, or [Fe/H]\,=\,$-$0.12\,$\pm$\,0.03, and
an age of 1.58\,$\pm$\,0.04 Gyr, where the errors quoted include
uncertainty in the main sequence fit, determined by the difference
between two adjacent isochrones. With these results, 25 stars out of
58 on the main sequence, or 43\%, have been identified as
photometric binaries.

\begin{figure}[pht]
\centerline{\psfig{figure=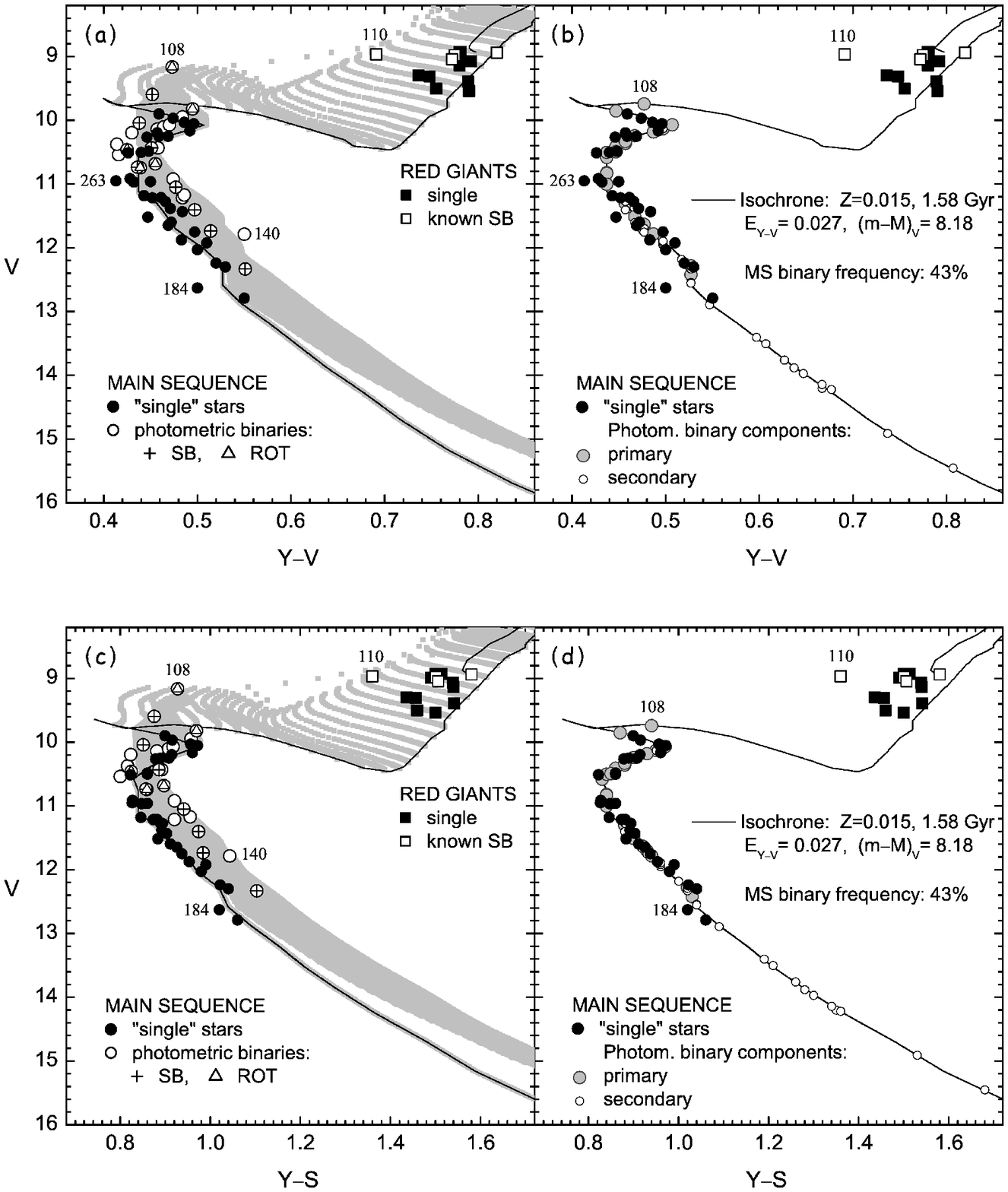,width=125truemm,angle=0,clip=}}
\captionb{4}{CMDs for NGC\,752, along with the results of the
least-squares minimization performed by varying the cluster
distance, metallicity and age.  Panels ($a$) and ($c$) display the
observed stars regarded photometrically as ``single" (dark circles)
and binaries (clear circles); the lightly shaded area denotes the
binary trajectories outlined by the best fit isochrone of 1.58 Gyr.
Panels ($b$) and ($d$) display the same ``single" stars (dark
circles) and the positions of photometric binary components (lightly
shaded and small open circles) on the same best fit isochrone.  The
theoretical isochrone (Padova models) in the observational plane of
the {\em Vilnius} system is from Bressan \& Tautvai{\v s}ien{\. e}
(1996).  A `kink' in the isochrone at {\em V}${\scriptstyle
\approx}$12.5 and {\em Y--V}${\scriptstyle \approx}$\,0.53 is an
artifact of the {\em T}$_{\rm eff}$-to-color transformations used in
their tabulations.  The gap in the two-star blend region along the
fainter end of the theoretical main sequence is a consequence of the
lower mass limit (0.6\,$M_\odot$) in the Padova models.  }
\end{figure}

Two CMDs with the best fit results, $V$,\,$Y$$-$$V$ and
$V$,\,$Y$$-$$S$, are shown in Figure~4. To visualize the impact of
the binary presence on the CMD, we show in the left-hand panels,
($a$) and ($c$), all possible trajectories of the model binaries
(the lightly shaded area). The appearance of a gap along and above
the fainter end of the theoretical main sequence is a natural
consequence of the lower mass limit (0.6\,$M_\odot$) in the Padova
models used. The cluster members regarded by our photometric
approach as single main-sequence stars and those identified as
binary candidates are shown by solid points and open circles,
respectively. The known spectroscopic binaries and rapid rotators
are marked with plus-signs and open triangles, respectively,
superimposed on symbols denoting photometric binary stars. The
right-hand panels, ($b$) and ($d$), show, for comparison purposes,
the same best fit isochrone, but with the model components of
photometric binaries plotted. The remaining stars, regarded as
``single", are also plotted. It is apparent from the picture that
the quantitative approach adopted here to finding binary candidates
is sufficiently fruitful and that unresolved binaries need not be
necessary excluded from the CMD fitting process.

The best isochrone fit to the observed main-sequence stars (Figure
4) gives the cluster age 1.6 Gyr, which is in agreement with the
latest values given in the literature. For example, Pols et al.
(1998) arrived at exactly the same age, using their own theoretical
isochrones with the same metallicity and reddening as in our case
but taking the distance modulus by 0.1 mag larger than our best-fit
value. Assuming the same distance modulus as in Pols et al. (1998),
$(m$$-$$M)_0$\,=\,8.20, the same age was also obtained by
Kozhurina-Platais et al. (1997) using the Yale isochrones and by
Anthony-Twarog \& Twarog (2004, 2006) using the isochrones by
Girardi et al. (2002) and VandenBerg et al. (2006), respectively
(all of the above mentioned isochrones were with convective core
overshooting). According to our best-fit isochrone of 1.6 Gyr, the
turn-off mass in NGC\,752 is $\sim$\,1.5\,$M_\odot$.

Although the 1.6 Gyr isochrone in Figure~4 fits the turn-off and the
extent of the main sequence quite well, the red clump giants fall
near, but not at the very base of the asymptotic giant branch. This
fact may be related to the presence of some systematic error in the
transformations of the Padova isochrones to {\em Vilnius} colors.
When we calculated metallicities [Fe/H] from the transformed {\em
Vilnius} colors along the entire isochrones concerned, we found that
[Fe/H] values for the model main-sequence stars were quite
consistent with the corresponding $Z$ values of the isochrones, but
the colors of model RGB stars gave [Fe/H] values systematically
lower (up to 0.1 dex) than for the main sequence. This is in the
opposite sense to what has been found in the case of the real
cluster stars, where the red clump giants were found to have
slightly higher photometric metallicities than the main-sequence
F-type stars (\S\,4.2).

Let us briefly consider the other two parameters of the cluster,
deduced through the minimization techniques, metallicity and
distance modulus. The best fit value of $Z$, corresponding to
[Fe/H]\,=\,$-$0.12\,$\pm$\,0.03, is only slightly higher than the
mean metallicity determined in \S\,4.2 for F-type stars,
[Fe/H]\,=\,--\,0.14\,$\pm$\,0.03 (s.d.), and both values agree within
the errors. However, this cannot be said about the cluster distance.
The best fit value, $(m$$-$$M)_V$=\,8.18, is by 0.2 mag lower than
that derived by averaging the individual distance moduli in \S\,4.1
(see Table~4) and by $\sim$\,0.1 mag lower than the one representing a
clearly defined peak of the $(m$$-$$M)_V$ distribution in Figure~3.
From our observations, an average modulus of $\sim$\,8.2 mag can be
attained only in the case of the red giants, if to apply
pre-Hipparcos calibrations of luminosity indices of the {\em
Vilnius} system. (For dwarf stars, pre- and post-Hipparcos
calibrations do not differ so much.) On the other hand, the value
obtained from the best fit is in good agreement with the results
based on the two eclipsing binaries (Milone \& Terrell 1996) or
obtained in a few of the photometric studies (see Table~4 in
\S\,4.1).

\begin{figure}[h!]
\centerline{\psfig{figure=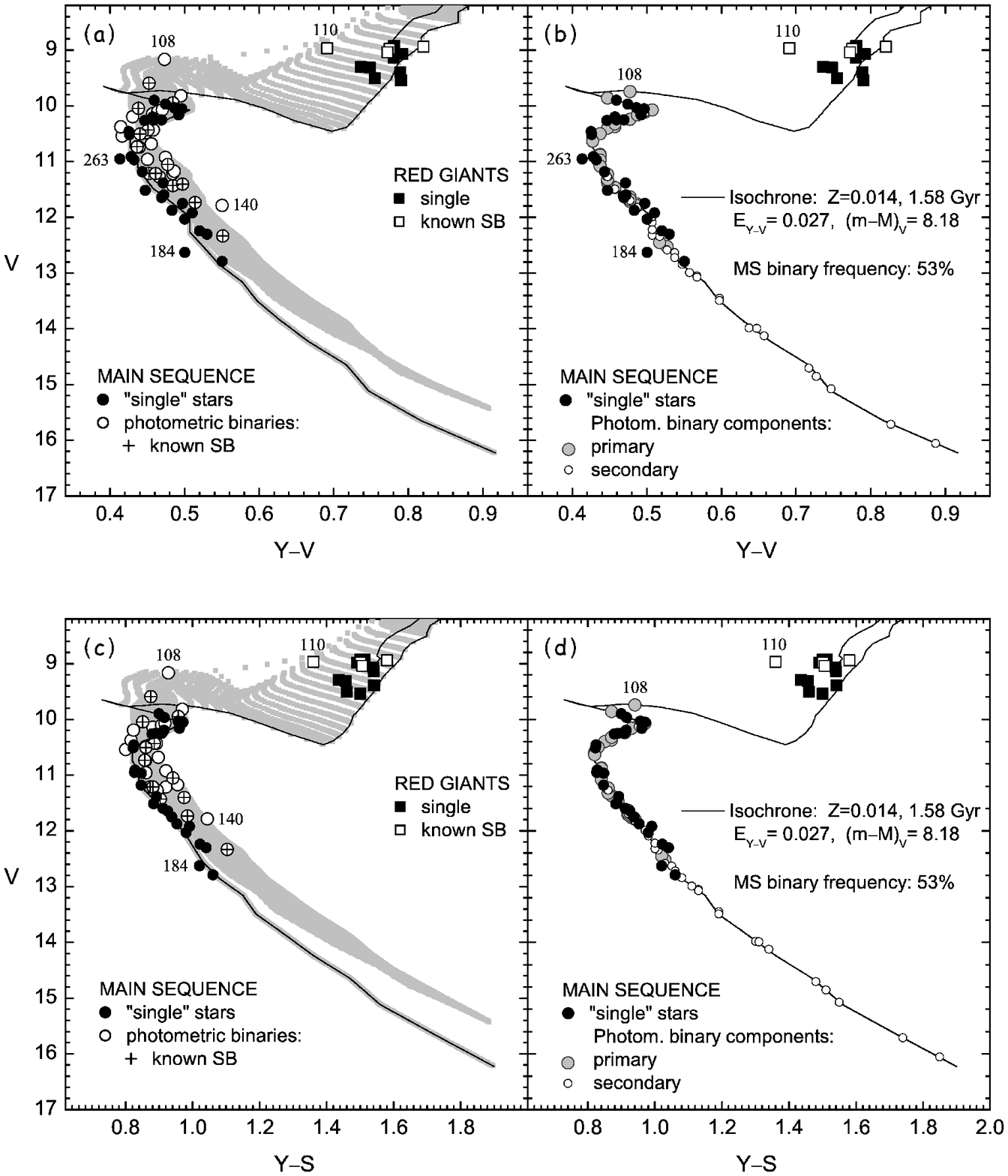,width=125truemm,angle=0,clip=}}
\captionb{5}{CMDs for NGC\,752, along with the results of the
least-squares minimization performed with a fixed metallicity
([Fe/H]\,=\,--0.14), but varying the cluster distance and age.
Notations are the same as in Figure~4.}
\end{figure}

\vskip -3pt

The least-squares minimization procedure has been also performed by
varying two of the cluster parameters instead of three. Figure~5
shows the results obtained with a fixed value of metallicity
([Fe/H]\,=\,--0.14, as derived for the main-sequence stars in
\S\,4.2), but the values of distance and age set free. In this case,
we arrive at exactly the same distance and age and in Figure~4, but
the fraction of photometric binaries gets larger: 53\% (31/58).

\begin{figure}[t!]
\centerline{\psfig{figure=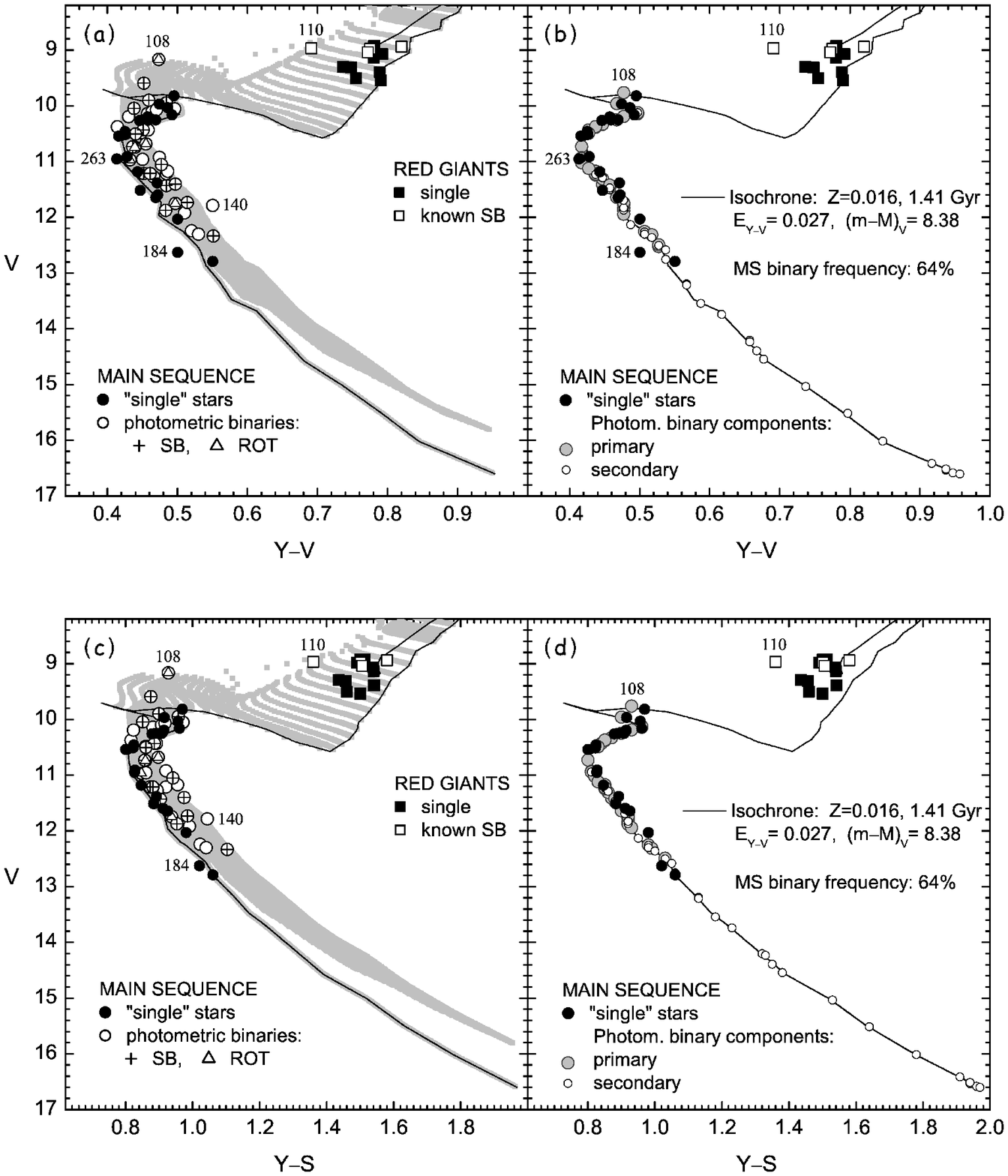,width=125truemm,angle=0,clip=}}
\captionb{6}{CMDs for NGC\,752, along with the results of the
least-squares minimization performed with a fixed distance modulus,
({\em m}--{\em M})$_V$=\,8.38 (Table~3), but varying the cluster
metallicity and age. Notations are the same as in Figure~4.}
\end{figure}

\vskip -3pt

In Figure~6 we show the results obtained with a fixed distance
modulus, $(m$$-$$M)_V$=\,8.38 (see Table 3 in \S\,4.1), but varying
[Fe/H] and age. The minimization yields the metallicity
[Fe/H]\,=\,$-$0.08 (or $Z$\,=\,0.016), an age of 1.4 Gyr, and the
binary fraction increased to 64\% (37/58). With this value of
distance modulus, the main clump of red giants is better reproduced
by the isochrone, but the ``single-star" main sequence seems likely
to be under-represented. We note, that the CMD fitting with both
[Fe/H] and $(m$$-$$M)_V$ fixed at the values derived from individual
stars yields no unique solution for the cluster age as the
``single-star" sequence becomes represented by too few stars.

To decide which of the three best-fit results, displayed in Figures
4, 5 and 6, describes the cluster's binary population better, we
used a formal quantitative parameter $s/N$, where $s$ is the overall
sum of the squared differences according to formula (5) and $N$ the
number of stars used in the fit ($N$\,=\,58). In the above three cases,
the minimization gave $s/N$ values of 0.0034, 0.0039 and 0.0046,
respectively. Therefore, we assumed as final best-fit results the
parameters from Figure~4, corresponding to the smallest $s/N$ value:
$(m$$-$$M)_V$=\,8.18\,$\pm$\,0.03 (or $(m$$-$$M)_0$=\,8.06),
[Fe/H]\,=\,$-$0.12\,$\pm$\,0.03, an age of 1.58\,$\pm$\,0.04 Gyr,
and a binary fraction of 43\%.

For 33 main-sequence stars, specified by our method as ``single"
(dark circles in Figure~4), the average distance modulus over the
individual values given in Table~2 is
$(m$$-$$M)_V$\,=\,8.25$\pm$0.05 (s.e.m.), i.e., closer to the
best-fit results than to the average value over 16 stars selected in
\S\,4.1 as representing the lower envelope of the main sequence.
According to Table~2, the mean reddening for the group of 33
``single" stars is $E_{Y-V}$=\,0.030\,$\pm$\,0.002 (s.e.m.) and the
mean metallicity [Fe/H]\,=\,$-$0.13\,$\pm$\,0.01 (only stars with
[Fe/H] determined from the color calibrations were considered).

\vskip2mm

\subsubsectionb{4.3.3}{Binary population}

In this subsection, we shall briefly discuss the population of the
25 main-sequence binaries, identified through our photometric
approach.

The frequency of such binaries, 43\% (25 stars out of 58), is close
to the upper limit of percentages evaluated from observations of
other open clusters by using different photometric approach
(15$-$40\%\,; see, e.g., Janes \& Kassis 1997). A higher frequency of
binaries (around 60$-$70\%) is, however, expected from cluster
simulations (Kroupa 1995) or from statistical analysis based on
photometric properties of binaries in such clusters as the Pleiades
(K{\"a}hler 1999). For comparison, the known spectroscopic binaries
in NGC\,752 constitute, according to Table~1 of DLMT, 37\% (22/60)
of the main-sequence members which were surveyed for radial-velocity
variability or, at least, have multiple radial velocity observations
reported.
\vskip3mm

\hbox{
\parbox{50mm}{\footnotesize\tabcolsep=3pt
\begin{tabular}{rrrrr|rrrrrr}
\multicolumn{10}{c}{\parbox{44mm}{\baselineskip=10pt {\smallbf\ \
Table~6.}{\small\ Masses for the 25 photometric binaries.}}}\\[6pt]
\tablerule\\[-11pt]
   &&&&&&&&&&\\[-6pt]
\multicolumn{1}{c}{No.} && \multicolumn{1}{c}{$M_1$} &
\multicolumn{1}{c}{$M_2$} &&& \multicolumn{1}{c}{~No.~} &&
\multicolumn{1}{c}{$M_1$} &
\multicolumn{1}{c}{$M_2$} \\[-1pt]
\tablerule\\[-18pt]
   &&&&&&&&&&\\
 10 && 1.59 &  1.21  &&& 140 && 1.15 &  1.05  \\
 12 && 1.67 &  1.22  &&& 187 && 1.47 &  1.25  \\
 41 && 1.69 &  1.28  &&& 192 && 1.47 &  0.95  \\
 61 && 1.58 &  1.33  &&& 193 && 1.60 &  1.11  \\
 63 && 1.26 &  1.22  &&& 214 && 1.56 &  0.86  \\
 64 && 1.53 &  0.86  &&& 217 && 1.56 &  0.89  \\
 66 && 1.28 &  1.26  &&& 218 && 1.65 &  1.11  \\
 74 && 1.47 &  0.97  &&& 234 && 1.43 &  1.17  \\
 88 && 1.25 &  0.72  &&& 235 && 1.30 &  0.90  \\
 96 && 1.58 &  0.92  &&& 237 && 1.13 &  0.78  \\
108 && 1.70 &  1.67  &&& 261 && 1.28 &  1.15  \\
126 && 1.63 &  1.14  &&& 300 && 1.70 &  1.36  \\
135 && 1.35 &  0.87  &&&     &&      &        \\[-2pt]
 \tablerule\\
\end{tabular}
}
\hspace{10mm}
\parbox{60mm}{
\psfig{figure=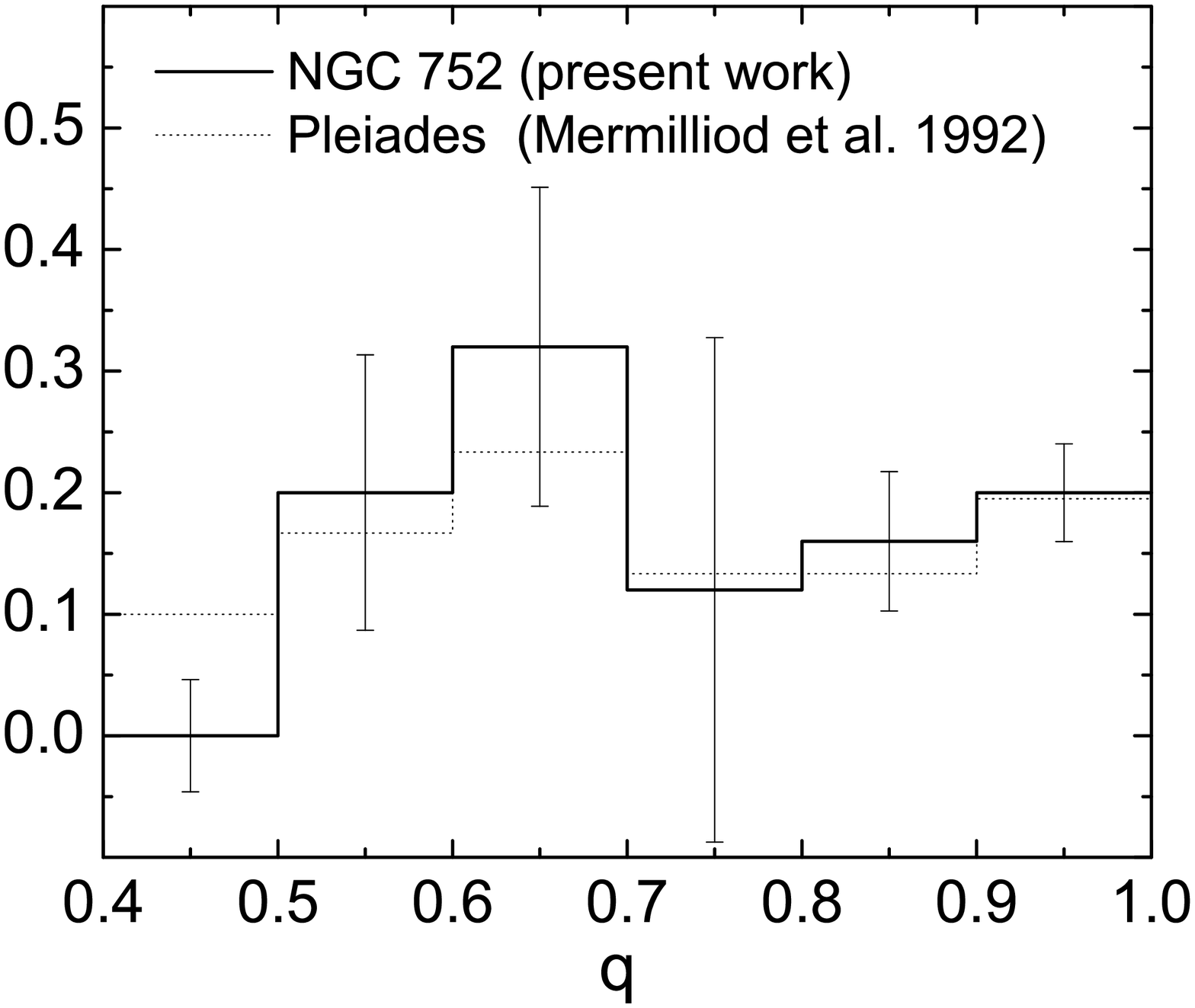,width=59truemm,angle=0,clip=0}
\captionb{7}{The mass ratio distribution for the
photometric binaries in NGC\,752 (heavy line with error bars). The
distribution for the Pleiades (Mermilliod et al. 1992) is reproduced
as a light dotted line.}
}
}
\vspace{2mm}

Because of different observational limitations and detection biases,
photometric and radial-velocity techniques do not necessary give
self-consistent results on binarity status. Out of 18 spectroscopic
binaries included in our photometric treatment, only half (or nine
stars) were identified photometrically as double. As one would
expect, most of these nine stars (6/9) are double-lined binaries
(SB2). The other half of the known spectroscopic binaries, of which
seven are single-lined binaries (SB1) and only two SB2s, fell into
the group of the 33 photometrically ``single" stars.

A list of the 25 binaries is given in Table~6, together with the
photometrically deconvolved masses of their components. From a
series of numerical simulations, we estimated that the primary and
secondary masses should be accurate to within 0.04\,$M_\odot$ and
0.08\,$M_\odot$, respectively.

In Figure~7 we show the distribution of mass ratio $q$\,=\,$M_2/M_1$
of the components, normalized by dividing by the number of binary
stars ($N$\,=\,25). The error bars represent 95\% confidence intervals
obtained after rescaling the standard error of each bin, estimated
from five numerical simulations, by a $t$-Student factor of 2.78.
Shown for comparison is the distribution from Mermilliod et al.
(1992) for 30 spectroscopic and photometric binaries of types F5--K0
in the Pleiades, reproduced from their Fig.~7.

It should be noted that binaries with mass ratios $q$\,$<$\,0.5 are
below the level of photometric detection ($\Delta m\approx$\,4 mag).
Therefore, a decline in the frequency of binaries, seen below
$q$\,$\approx$\,0.6, may be induced by this limit. For
$q$\,$>$\,0.6, the distribution in general resembles that given by
Mermilliod et al. (1992) for the Pleiades cluster. There is little
evidence in Figure~7 of bimodality of the distribution with a second
peak near $q$\,=1, as expected from a statistical analysis by
K{\"a}hler (1999) of the Pleiades binaries deconvolved using
synthetic clusters. The size of the error bar for the bin at
$q$\,$\approx$\,0.75 is too large to consider the dip at this mass
ratio real and hence to admit that the distribution is indeed rising
toward $q$\,=1.

\vskip2mm

\sectionb{5}{SUMMARY AND CONCLUSIONS}

New photoelectric seven-color observations in the {\em Vilnius}
system for 62 members of the open cluster NGC\,752 were used to
derive the main cluster parameters: foreground reddening, distance,
metallicity and age. The apparent distance modulus, based on
individual absolute magnitudes of 10 single RGB stars, is found to
be $(m$$-$$M)_V$\,=\,8.40$\pm$0.05, which is consistent with that
derived for 16 F-type stars lying on the lower envelope of the main
sequence, 8.37$\pm$0.04 mag (the errors quoted in this chapter are
the standard errors of the mean, therefore we give also the number
of stars used in averaging). The reddening to the cluster, derived
from both the red giants and F-type stars ($N$\,=\,26), is
$E_{Y-V}$\,=\,0.027\,$\pm$\,0.002, or $E_{B-V}$\,=\,0.034. With the
above values from individual stars, we obtain a true distance
modulus $(m$$-$$M)_0$\,=\,8.27$\pm$0.03, or a distance to the
cluster of 450$\pm$10 pc.

The mean metallicity determined for the 10 red giants concentrated
in the region of the clump, [Fe/H]\,=\,$-$0.08 $\pm$\,0.03, appears
to be slightly higher (formally at the 2$\sigma$ level) than that
found for F-type stars of the lower envelope of the main sequence,
[Fe/H]\,=\,$-$0.14\,$\pm$\,0.01 ($N$\,=\,8). This is consistent with a
similar result obtained for NGC\,7789 by Barta\v si\= ut\. e \&
Tautvai{\v s}ien{\. e} (2004) from comparison of photometric
metallicities between the first ascent RGB stars and the more
evolved red giants in the clump. These findings received an
explanation in terms of evolutionary changes in the intensities of
carbon and nitrogen molecular bands many of which fall in the {\em
Vilnius} passbands used for the determination of [Fe/H]. Therefore,
a suggestion can be made that giants in later evolutionary phases
may not have zero-age surface values of metallicity.

We made use of the seven color data to identify the binary
population of the cluster, responsible for much of the scatter in
the main sequence. Based on the Padova theoretical models, we have
created random samples of artificial stars of various masses for
given metallicities and ages relevant to the cluster. Follow-up
application of the least-squares minimization algorithm allowed us
to find the best match between all the seven magnitudes and colors
of the observed stars and those of model single and binary stars by
varying in small steps the cluster distance, metallicity and age. In
this way, we arrived at the following best-fit parameters for the
cluster: $(m$$-$$M)_0$=\,8.06\,$\pm$\,0.03, or a distance of 410 pc,
[Fe/H]\,=\,$-$0.12\,$\pm$\,0.03, an age of 1.58\,$\pm$\,0.04 Gyr and
a binary fraction of 43\%.

We noted that the value of distance modulus determined from
individual stars and the one derived by the CMD fitting techniques
differ by as much as 0.2 mag. Given that the averages over
individual $(m$$-$$M)_V$ values for different groups of stars differ
substantially (Table~3), there is no reason to expect that absolute
magnitudes derived for individual stars would give a more accurate
distance than the CMD fit.

Finally, we conclude that our approach to fitting the CMD allows a
reliable determination of cluster parameters, and it seems to work
successfully for a cluster with the rich population of binaries.
However, for a more robust application of the method, a significant
improvement in the transformations of model luminosities, gravities
and temperatures to the {\em Vilnius} colors is required.

In the near future, CCD photometry in the {\em Vilnius} system will be
obtained in the region of the cluster down to the fainter members
which are still limited to photographic data. The stars observed
photoelectrically in the present work are to be used as standards in
the CCD reductions. The seven-color CCD data will hopefully allow us
to define the lower main sequence and hence to discuss possible
dissolution of the cluster from the low mass end up.

\vskip 5pt

ACKNOWLEDGMENTS.  We are thankful to A.\,G.\,Davis Philip and Imants
Platais for helpful comments on a draft of this paper.  This work was
supported by National Science Council of Taiwan and the Ministry of
Education and Science of Lithuania.

\goodbreak

\References

\refb Anthony-Twarog~B.\,J., Twarog~B.A. 2004, AJ, 127, 1000

\refb Anthony-Twarog~B.\,J., Twarog~B.A. 2006, PASP, 118, 358

\refb Arp~H. 1962, ApJ, 136, 66

\refb Barta\v si\= ut\. e~S., Tautvai{\v s}ien{\. e}~G. 2004, A\&SS,
294, 225

\refb Bartkevi\v cius~A., Sperauskas~J. 1983, Bull. Vilnius Obs.,
63, 3

\refb Bell~R.\,A. 1972, MNRAS, 157, 147

\refb Bressan~A., Tautvai{\v s}ien{\. e}~G. 1996, Baltic Astronomy,
5, 239

\refb Bressan~A., Fagotto~F., Bertelli~G., Chiosi~C. 1993, A\&AS,
100, 647

\refb Cannon~R.\,D. 1968, Ph. D. Thesis, University of Cambrige

\refb Cannon~R.\,D. 1970, MNRAS, 150, 111

\refb Canterna~R., Geisler~D., Harris~H.\,C., et al. 1986, AJ, 92,
79

\refb Crawford~D.\,L., Barnes~J.\,V. 1970, AJ, 75, 946

\refb Daniel~S.\,A., Latham~D.\,W., Mathieu~R.D., Twarog~B.A. 1994,
PASP, 106, 281 (DLMT)

\refb Dz{\`e}rv{\`i}tis~U., Paupers~O. 1993, A\&SS, 199, 77 (DP)

\refb Dz{\`e}rv{\`i}tis~U., Paupers~O. 1994, Baltic Astronomy, 3,
335

\refb Ebbighausen~E.\,G. 1939, ApJ, 89, 431

\refb Eggen~O.\,J. 1963, ApJ, 138, 356

\refb Eggen~O.\,J., Iben~I.,\,Jr. 1988, AJ, 96, 635

\refb Eggen~O.\,J., Sandage~A.\,R. 1964, ApJ, 140, 130

\refb Fagotto~F., Bressan~A., Bertelli~G., Chiosi~C. 1994, A\&AS,
105, 29

\refb Francic~S.\,P. 1989, AJ, 98, 888

\refb Friel~E.\,D., Janes~K.\,A. 1993, A\&A, 267, 75

\refb Friel~E.\,D., Janes~K.\,A., Tavarez~M., Scott~J., et al. 2002,
AJ, 124, 2693

\refb Garrison~R.\,F. 1972, ApJ, 177, 653

\refb Geisler~D. 1984, ApJ, 287, L85

\refb Gilroy~K.\,K. 1989, ApJ, 347, 835

\refb Gilroy~K.\,K., Brown~J.\,A. 1991, ApJ, 371, 578

\refb Girardi~L., Mermilliod~J.-C., Carraro~G. 2000, A\&A, 354,
892

\refb Girardi~L., Bertelli~G., Bressan~A., Chiosi~C., et al. 2002,
A\&A, 391, 195

\refb Gunn~J.\,E., Kraft~R.\,P. 1963, ApJ, 137, 301

\refb Hardy~E. 1979, AJ, 84, 319

\refb Heinemann~K. 1926, AN, 227, 193

\refb Hirshfeld~A., McClure~R.\,D., Twarog~B.\,A. 1978, in {\em The
HR Diagram}, eds.\break A.\,G.\,D.~Philip \& D.\,S.~Hayes, p.~163

\refb Hobbs~L.\,M., Pilachowski~C. 1986, ApJ, 309, 17L

\refb Hobbs~L.\,M., Thorburn~J.\,A. 1992, AJ, 104, 669

\refb Janes~K.\,A., Kassis~M. 1997, in {\em The 3rd Pacific Rim
Conference on Recent Development of Binary Star Research}, ed.
K.-C.\,\,Leung, ASP\,\,Conf.\,\,Ser., 130, 107

\refb Jennens~P.\,A., Helfer~H.\,L. 1975, MNRAS, 172, 681

\refb Johnson~H.\,L. 1953, ApJ, 117, 356

\refb Joner~M.\,D., Taylor~B.\,J. 1995, PASP, 107, 351

\refb K{\"a}hler~H. 1999, A\&A, 346, 67

\refb Kazlauskas~A. 1996, Baltic Astronomy, 5, 319

\refb Komarov~N.\,S., Shcherback~A.\,N. 1980, AZh, 57, 557

\refb Kozhurina-Platais~V., Demarque~P., Platais~I., et al. 1997,
AJ, 113, 1045

\refb Kroupa~P. 1995, MNRAS, 277, 1522

\refb Maeder~A. 1974, A\&A, 32, 177

\refb Mermilliod~J.-C., Rosvick~J.\,M., Duguennoy~A., Mayor~M. 1992,
A\&A, 265, 513

\refb Mermilliod~J.-C., Mathieu~R.\,D., Latham~D.\,W., Mayor~M.
1998, A\&A, 339, 423

\refb Meynet~G., Mermilliod~J.-C., Maeder~A. 1993, A\&AS, 98, 477

\refb Milone~E.\,F., Stagg~C.\,R., Sugars~B.\,A., McVean~J.\,R., et
al. 1995, AJ, 109, 359

\refb Milone~E.\,F., Terrell~D. 1996, in {\em The Origins,
Evolutions, and Destinies of Binary Stars in Clusters}, eds.
E.\,F.~Milone \& J.-C.~Mermilliod, ASP Conf. Series, 90, 283

\refb Nicolet~B. 1981a, A\&A, 97, 85

\refb Nicolet~B. 1981b, A\&A, 104, 185

\refb Nissen~P.-E. 1988, A\&A, 199, 146

\refb Pilachowski~C.\,A., Saha~A., Hobbs~L.\,M. 1988, PASP, 100,
474

\refb Platais~I. 1991, A\&AS, 87, 69

\refb Pols~O.\,R., Schr{\"o}der~K.-P., Hurley~J.\,R., Tout~C.\,A.,
Eggleton~P.\,P. 1998, MNRAS, 298, 525

\refb Rohlfs~K., Van{\'y}sek~V. 1962, Abh. Hamburg Sternw., 5, 341

\refb Roman~N.\,G. 1955, ApJ, 121, 454

\refb Romani~R.\,W., Weinberg~M.\,D. 1991, ApJ, 372, 487

\refb Roxburgh~I.\,W., Strittmatter~P.A. 1965, Z. Astrophys., 63, 15

\refb Rufener~F.\,G. 1988, {\em Catalogue of Stars Measured in the
Geneva Photometric System}, 4th ed., Observatoire de Geneve (CDS
Catalogue II/169)

\refb Schiller~S.\,J., Milone~E.\,F. 1988, AJ, 95, 1466

\refb Sestito~P., Randich~S., Pallavicini~R. 2004, A\&A, 426, 809

\refb Strai{\v z}ys~V. 1992, {\em Multicolor Stellar Photometry},
Pachart Publishing House, Tucson, Arizona

\refb Tautvai{\v s}ien{\. e}~G., Edvardsson~B., Barta\v si\= ut\.
e~S. 2003, Baltic Astronomy, 12, 532

\refb Twarog~B.\,A. 1983, ApJ, 267, 207

\refb Twarog~B.\,A., Ashman~K.\,M., Anthony-Twarog~B.\,J. 1997, AJ,
114, 2556

\refb van den Berg~M., Verbunt~F. 2001, A\&A, 375, 387

\refb VandenBerg~D.\,A., Bergbusch~P.\,A., Dowler~P.\,D. 2006, ApJS,
162, 375

\refb Wallerstein~G., Conti~P. 1964, ApJ, 140, 858

\refb Zdanavi{\v c}ius~K. 1996, Baltic Astronomy, 5, 549

\refb Zdanavi{\v c}ius~K., {\v C}ernien{\. e}~E. 1985, Bull.
Vilnius Obs., 69, 3

\refb {\v Z}itkevi{\v c}ius~V., Strai{\v z}ys~V. 1972, Bull. Vilnius
Obs., 34, 30

\vskip 10pt

{\em Note added in proof.} In the most recent work on NGC\,752 by Taylor
(2007, AJ, 134, 934), the reddening to the cluster has been reanalyzed
on the basis of all published estimates.  It has been stressed that
reddening values derived from K giants can suffer from systematic
effects, and the value based solely on F-type stars,
$E_{B-V}$\,=0.044$\pm$0.003, was adopted for the cluster.  In connection
with Taylor's work, we note that the mean reddening found in our work
for K giants, $E_{B-V}$=\,0.030\,$\pm$\,0.004, shows no indication of
greater than 1$\sigma$ difference from that for F-type stars considered
as single, i.e., $E_{B-V}$=\,0.035\,$\pm$\,0.003 (see Table~3).

\end{document}